\documentclass[aps,prl,reprint,amsmath,amssymb,superscriptaddress,noeprint,nofootinbib]{revtex4-2}
 
\pdfoutput=1

\usepackage{graphicx}
\usepackage{dcolumn}
\usepackage{bm}
\usepackage[version=3]{mhchem}
\usepackage{mathtools}
\PassOptionsToPackage{hyphens}{url}

\usepackage{float}
\usepackage{color}
\usepackage{graphicx}
\usepackage{amsmath,amssymb}
\usepackage{upgreek}
\usepackage{psfrag} 
\usepackage{graphicx}
\usepackage{braket}
\usepackage{units}
\usepackage{times}
\usepackage{lineno}
\usepackage[svgnames]{xcolor}
\definecolor{myColor}{rgb}{0.02,0.12,0.3}
\definecolor{myciteColor}{rgb}{0.39,0.7,0.89}
\usepackage[colorlinks=true,allcolors=blue]{hyperref}
\newcommand{\ext}{\text{ext}}
\newcommand{\coll}{\text{coll}}
\newcommand{\FD}{\text{FD}}
\newcommand{\eq}{\text{eq}}
\newcommand{\F}{\mathrm{F}}

\newcommand{\dd }{\mathrm{d}}
\newcommand{\kF}{k_\mathrm{F}}
\newcommand{\kFa}{k_\mathrm{F}a_\mathrm{s}}
\newcommand{\EF}{E_\mathrm{F}}
\newcommand{\vF}{v_\mathrm{F}}

\newcommand{\diff}{\mathrm{d}}
\newcommand{\TF}{T_\mathrm{F}}
\newcommand{\as}{a_\mathrm{s}}
\newcommand{\bas}{\bar{a}_\mathrm{s}}
\newcommand{\bc}{\bar{c}}
\newcommand{\kB}{k_\mathrm{B}}
\renewcommand{\a}{\mathrm{a}}
\newcommand{\s}{\mathrm{s}}
\newcommand{\ws}{\omega_\mathrm{s}}
\newcommand{\Us}{U_\mathrm{s}}
\newcommand{\Bk}{\mathbf{k}}
\newcommand{\Br}{\mathbf{r}}
\newcommand{\Bp}{\mathbf{p}}
\newcommand{\Bq}{\mathbf{q}}
\newcommand{\p}{\mathrm{p}}
\newcommand{\R}{\mathrm{R}}
\newcommand{\Vol}{\mathcal{V}}
\newcommand{\tchi}{\tilde{\chi}}
\newcommand{\grid}{\text{grid}}

\newcommand{\be}{\begin{equation}}
\newcommand{\ee}{\end{equation}}
\newcommand{\bea}{\begin{eqnarray}}
\newcommand{\eea}{\end{eqnarray}}

\newcommand{\meanv}[1]{\langle #1 \rangle}

\newcommand{\bb}[1]{\left( #1 \right)}

\newcommand{\bbcro}[1]{\left[ #1 \right]}

\newcommand{\rF}{\mathrm{F}}
\newcommand{\ii}{i}
\newcommand{\eee}{e}
\newcommand{\rr}{\mathbf{r}}
\newcommand{\rL}{\mathrm{L}}

\newcommand{\qq}{\mathbf{q}}

\newcommand{\pp}{\mathbf{p}}

\renewcommand{\Re}{\mathrm{Re}}
\renewcommand{\Im}{\mathrm{Im}}

\makeatletter
\def\maketitle{
\@author@finish
\title@column\titleblock@produce
\suppressfloats[t]}
\begin{document}

\title{Emergence of Sound in a Tunable Fermi Fluid}

 \author{Songtao Huang$^{*\dagger}$}
 \affiliation{Department of Physics, Yale University, New Haven, Connecticut 06520, USA}
 \author{Yunpeng Ji$^{*}$}
 \affiliation{Department of Physics, Yale University, New Haven, Connecticut 06520, USA}
 \author{Thomas Repplinger}
 \affiliation{Laboratoire de Physique Théorique,
Université de Toulouse, CNRS, UPS, 31400, Toulouse, France}
 \author{Gabriel G. T. Assump{\c{c}}{\~a}o}
 \affiliation{Department of Physics, Yale University, New Haven, Connecticut 06520, USA}
 \author{Jianyi Chen}
 \affiliation{Department of Physics, Yale University, New Haven, Connecticut 06520, USA}
 \author{Grant L. Schumacher}
 \affiliation{Department of Physics, Yale University, New Haven, Connecticut 06520, USA} 
\author{Franklin J. Vivanco}
 \affiliation{Department of Physics, Yale University, New Haven, Connecticut 06520, USA}
 \author{Hadrien Kurkjian$^{\dagger}$}
 \affiliation{Laboratoire de Physique Théorique,
Université de Toulouse, CNRS, UPS, 31400, Toulouse, France}
\affiliation{Laboratoire de Physique Théorique de la Matière Condensée,
Sorbonne Université, CNRS, 75005, Paris, France}
 \author{Nir Navon}
 \affiliation{Department of Physics, Yale University, New Haven, Connecticut 06520, USA}
 \affiliation{Yale Quantum Institute, Yale University, New Haven, Connecticut 06520, USA}

\def\thefootnote{*}\footnotetext{These authors contributed equally to this work}
\def\thefootnote{$\dagger$}\footnotetext{songtao.huang@yale.edu, hadrien.kurkjian@cnrs.fr}

\date{\today}

\begin{abstract}

Landau's Fermi-liquid (FL) theory has been successful at the phenomenological description of the normal phase of many different Fermi systems. Using a dilute atomic Fermi fluid with tunable interactions, we investigate the microscopic basis of Landau's theory with a system describable from first principles. We study transport properties of an interacting Fermi gas by measuring its density response to a periodic external perturbation. In an ideal Fermi gas, we measure for the first time the celebrated Lindhard function. As the system is brought from the collisionless to the hydrodynamic regime, we observe the emergence of sound, and find that the experimental observations are quantitatively understood with a first-principle transport equation for the FL. When the system is more strongly interacting, we find deviations from such predictions. Finally, we observe the shape of the quasiparticle excitations directly from momentum-space tomography and see how it evolves from the collisionless to the collisional regime. 
Our study establishes this system as a clean platform for studying Landau's theory of the FL and paves the way for extending the theory to more exotic conditions, such as nonlinear dynamics and FLs with strong correlations in versatile settings.

\end{abstract}
\maketitle

\begin{figure*}[!hbt]
  \centering 
  \includegraphics[width=2\columnwidth]{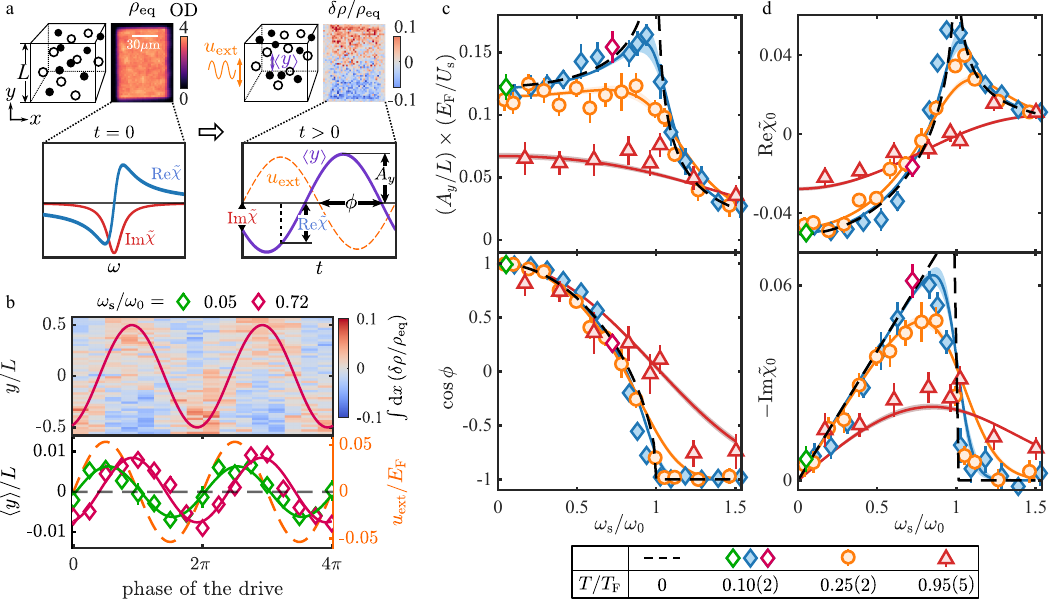}
  \caption{ \textbf{Probing the density response function $\tilde{\chi}$ and measurement of the Lindhard response function.} (\textbf{a}) Top panel: sketch of the experimental protocol. A homogeneous spin-1/2 Fermi gas of $^6$Li atoms is trapped in a cuboidal optical box with height $L$ along the shaking axis $y$. The black and white circles represent the two spin states. Note that gravity's direction is also along the $y$ axis and compensated for with a magnetic gradient. (Left) \emph{In situ} optical density (OD) of the gas in equilibrium with a density $\rho_\text{eq}$. (Right) The perturbation $u_\mathrm{ext}$ is turned on for variable time $t$. The purple double-arrow shows the instantaneous CoM $\braket{y}$. The \emph{in situ} OD of the density difference is shown, where $\delta\rho=\rho-\rho_\text{eq}$. Bottom panel: (Left) sketch of the real (blue solid line) and imaginary (red solid line) parts of an example response function $\tchi$ in the frequency domain. (Right) The resulting CoM response (solid purple line) under the drive (orange dashed line). $\Im\tchi$ ($\Re\tchi$) corresponds to the quadrature (in phase) response to the drive. $A_y$ and $\phi$ are the response amplitude and relative phase to the drive, respectively. 
  (\textbf{b}) Top panel: A heat map showing the integrated density variation in the shaking direction for $\ws/\omega_0\approx0.72$ at various phases of the drive. The red line is a guide to the peaked density variation. Bottom panel: The extracted CoM (left axis) from the top panel and the drive (right axis). We extract $A_y$ and $\phi$ by fitting a sinusoidal function (red and green solid lines) to the CoM. The error bars are statistical uncertainties from at least four individual measurements for each point and are smaller than the marker size. Different colors correspond to different drive frequencies (see the legend).
  (\textbf{c}) $A_y$ (top) and $\phi$ (bottom) extracted from the density response of a non-interacting Fermi gas at three different temperatures (see the legend). 
  (\textbf{d}) $\Re\tchi_0$ (top) and $\Im\tchi_0$ (bottom) of the response function extracted from (c) using Eq.~\eqref{eqn: tchi}. The error bars in (c) and (d) are the uncertainties from the fit. The solid lines in (c) and (d) are calculated from the linear response theory (see text); the bands include the uncertainties on the temperature determinations of the experiment.
  \label{FIG:1}}
\end{figure*}

\begingroup
\renewcommand{\addcontentsline}[3]{}% Remove functionality of \addcontentsline
\section{I. Introduction}
Landau's Fermi-liquid (FL) theory, developed in 1956~\cite{Landau1956}, is a paradigm at the heart of our understanding of many interacting Fermi systems~\cite{NozieresPines1966}. 
This theory relies on the argument that a smooth connection exists between the ideal Fermi gas and the interacting system as interactions are slowly turned on from zero. The effect of interactions is to adiabatically map the non-interacting particles onto quasiparticles: the interacting Fermi system can then be understood as a gas of quasiparticles having a phase-space distribution that equals the one for the non-interacting gas, with renormalized properties and with emergent collective properties~\cite{baym2008landau}.

This framework significantly simplifies the description of complicated many-body problems, encapsulating microscopic physics into a small number of phenomenological parameters, the so-called Landau parameters. Landau's theory has been successful in explaining both equilibrium and near-equilibrium properties of numerous Fermi systems, such as liquid $^3$He~\cite{ketterson1976physics}, normal metals~\cite{PhysRev.143.133}, heavy fermion materials~\cite{stewart1984heavy}, nuclear matter~\cite{larionov1999zero}, and twisted bilayer graphene~\cite{jaoui2022quantum}.

However, Landau's argument has remained theoretical and unexamined experimentally since `real-world' systems typically do not permit the continuous tuning of interactions. Consequently, none of those systems to which Landau's theory is applicable can be continuously connected to the original non-interacting state; the Landau parameters characterizing the Fermi liquid must then be determined phenomenologically (with some notable exceptions~\cite{PhysRevB.105.235107,  PhysRev.187.371,backman1973calculation,PhysRevB.50.1684}). The unsolved connections of Landau's theory to microscopics have important implications. For example, there are long-standing discrepancies between theory and experiments on the transport coefficients of $^3$He~\cite{baym2008landau}.

\section{II. Experimental protocol}

Ultracold Fermi gases provide a textbook platform for studying FL physics~\cite{PhysRevLett.102.230402,nascimbene2009collective,scazza2017repulsive}. First, in the dilute limit, they accurately implement the Hamiltonian of fermions interacting with contact interactions - characterized by the s-wave scattering length $\as$. This model Hamiltonian has played a central role in quantum many-body physics~\cite{lee1957many,huang1957quantum,giorgini2008theory}. One of its outstanding properties is that it is amenable to a FL description in the weak interaction limit, where the Landau parameters can be calculated analytically \emph{ab initio}~\cite{galitskii1958energy,abrikosov1958concerning,lifshitz2013statistical}. Secondly, the tunability of $\as$~\cite{RevModPhys.82.1225} provides a physical realization of the theoretical idea of the switching on of interactions. 

Here we use a uniform quantum gas of $^6$Li fermions to study how sound emerges as interactions are gradually turned on. In the non-interacting limit, this system does not possess any sound mode. As interactions increase, the excitations of this Fermi system evolve from broad particle-hole excitations (in the collisionless regime) to a narrow collective sound mode (in the hydrodynamic one). For weak interactions, our data are in excellent agreement with a quasiparticle transport equation derived from first principles. Furthermore, our measurements of the sound attenuation agree with an exact yet previously untested prediction~\cite{jensen1968exact,brooker1968transport}.

Our experiment starts with a spatially uniform quantum gas of spin-1/2 fermions, in a balanced mixture of two internal (spin) states labeled $\ket{\uparrow}$ and $\ket{\downarrow}$ (for more details, see~\cite{ji2022stability,ji2024observation}). The atoms are confined in an optical cuboidal box, see Fig.~\ref{FIG:1}a. Throughout our measurements, we vary the density of the gas so that the Fermi energy $\EF=\frac{\hbar^2}{2m}(6\pi^2 \rho_\text{eq})^{2/3}$ ranges from $\approx 2\pi\hbar\times \unit[3]{kHz}$ to $\approx 2\pi\hbar\times \unit[10]{kHz}$ (where $\rho_\text{eq}$ is the equilibrium density of one spin population and $m$ is the mass of the $^6$Li atom). We control $\as$ via a magnetic Feshbach resonance (see Appendix A). This allows us to explore the parameter space of $0 \lesssim |\kFa| \lesssim  \infty$ and $0.1 \lesssim T/\TF \lesssim 0.95$, where $T$ is the temperature of the gas, $\TF$ is the Fermi temperature, and $\kF$ is the Fermi wavevector. Within this parameter space, we focus our study on the normal phase, \emph{i.e.}, $T>T_\text{c}$ where $T_\text{c}$ is the superfluid transition temperature.

\section{III. Density response function and measurement of the Lindhard function}
After preparing an equilibrium Fermi gas at the desired interaction strength, we turn on a spatially uniform shaking force along the $y$ axis. The driving potential is $u_\mathrm{ext}(\Br,t)=-(U_\s/L) y \sin(\ws t)$, where $L$ is the box height along the $y$ axis, $U_\s$ is the drive amplitude, and $\omega_\s$ is the drive frequency (see Fig.~\ref{FIG:1}a). This gradient potential predominately couples to the longest wavelength excitations in momentum space~\cite{supp}, \emph{i.e.}, with a wavevector along the $y$ axis $q_\mathrm{ext}=|\Bq_\mathrm{ext}|\equiv\pi/L$ that ranges between $\approx 0.01-0.02\times\kF$; the typical frequency scale of these excitations is $\omega_0\equiv q_\mathrm{ext}\vF$, where $\vF$ is the Fermi velocity. Subsequently, we measure the center of mass (CoM) of the gas, $\braket{y}\propto \int \diff \Br\,\rho(\Br,t)y$ where $\rho(\Br,t)$ is the density distribution of one of the spin components of the shaken gas; since the perturbation $u_\text{ext}$ is spin independent, we only consider one of the spins in this work. The CoM reaches a steady state that oscillates at the same frequency $\ws$ as the drive after some shaking time (typically $\unit[500]{ms}$). We then extract the oscillation amplitude $A_y$ and the phase $\phi$ relative to the drive using a sinusoidal fit of fixed frequency $\omega_\s$ (see Fig.~\ref{FIG:1}b and Appendix B). 

\begin{figure*}[!hbt]
  \centering
  \includegraphics[width=2\columnwidth]{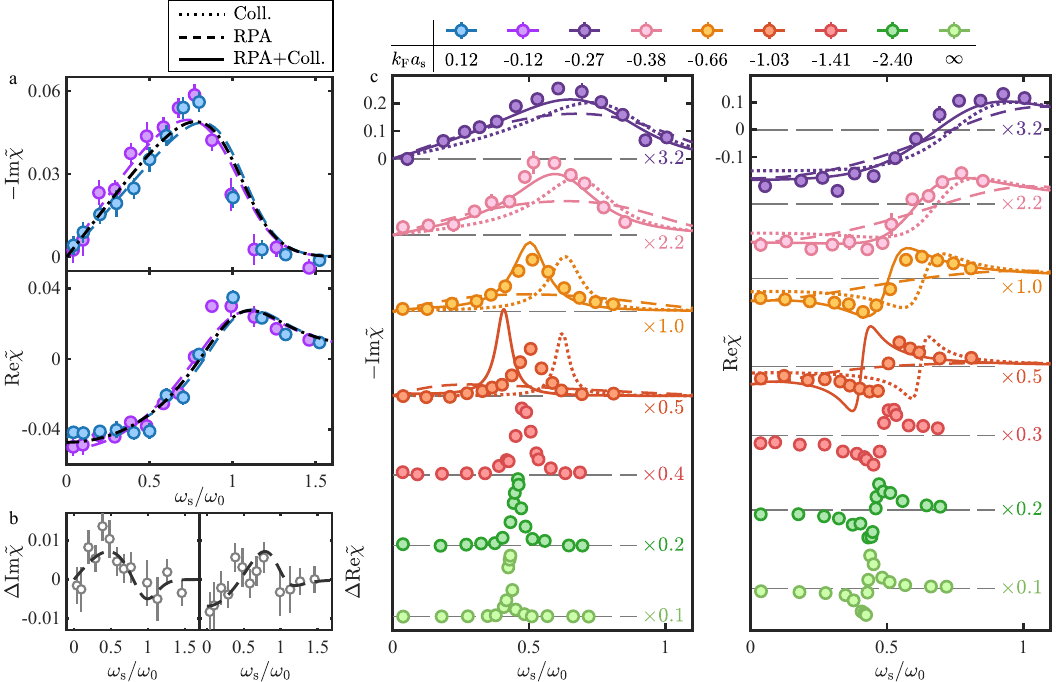}
  
  \caption{\textbf{Emergence of sound in a tunable Fermi liquid.} (\textbf{a}) Effect of mean-field interactions in the response of a weakly interacting Fermi gas, $\Im\tchi$ (top) and $\Re\tchi$ (bottom) for $\kFa=\pm 0.12$ (blue and purple points). The theoretical predictions from the RPA are shown as dashed lines, the ideal Lindhard response as black dot-dashed line, see text; $\EF\approx2\pi\hbar\times\unit[6]{kHz}$ and $T/\TF\approx0.2$. (\textbf{b}) Difference of the corresponding $\Im\tchi$ (left) and $\Re\tchi$ (right) for $\kFa= 0.12$ and $-0.12$, respectively; dashed lines are the predictions from the RPA. (\textbf{c}) Emergence of first sound and breakdown of a first-principle kinetic description. Theoretical predictions are shown: RPA alone (dashed), collisions alone (dotted) and the full numerical solution of Eq.~\eqref{eq: transport equation} (solid). The gray long dashed lines are the corresponding zeroes for each spectrum; a scaling factor is shown on the right.
  \label{FIG:2}}
\end{figure*}

In the experiment, we adjust the drive strength $U_\s/E_\F$ so that the response amplitude $A_y\propto U_\s$, \emph{i.e.}, the system is within the linear response regime. We relate the (dimensionless) density response function $\tchi(\omega)\equiv \delta\Tilde{\rho}(\mathbf{q}_\mathrm{ext},\omega)/U_\s\times (2\EF/\kF^3)$ to the CoM response using linear response theory~\cite{NozieresPines1966} (see Fig.~\ref{FIG:1}a and~\cite{supp}): 
\begin{align}
    \tchi&=-\frac{\pi^2}{24}\frac{\EF}{\Us}\frac{A_y}{L}e^{-i\phi}\label{eqn: tchi},
\end{align}
where $ \delta\tilde{\rho}(\Bq,\omega)\equiv \int\diff\Br\diff t~ \delta\rho(\Br,t) e^{-i(\Bq\cdot\Br-\omega t)}$ is the Fourier transform of the density deviation from the equilibrium $\delta\rho=\rho-\rho_\text{eq}$; from now on, we will use the tilde to denote the Fourier transform, and the prefix $\delta$ to denote the deviation from equilibrium. The real and imaginary parts of $\tchi$ are therefore proportional to the in-phase response ($\ws t=\mathcal{N} \pi/2$ where $\mathcal{N}$ is a positive integer) and the quadrature response ($\ws t=\mathcal{N} \pi$), respectively. In the field of condensed matter, the density response function is also known as the dielectric function, which gives access to the conductivity~\cite{girvin2019modern,anderson2019conductivity,wu2015probing}. 

We benchmark our experiment protocol by measuring the density response of an ideal Fermi gas. In Fig.~\ref{FIG:1}c, we show $A_y$ and $\cos\phi$ as a function of $\omega_\s/\omega_0$ for various $T/\TF$; in Fig.~\ref{FIG:1}d, we show the corresponding response functions $\tchi_0$. The theoretical prediction $\tchi_{\rm L}$, derived by Lindhard in 1954~\cite{lindhard1954properties}, is a classic result in condensed matter physics~\cite{girvin2019modern}. It is shown as solid lines in Fig.~\ref{FIG:1}c-d and is in excellent agreement with the experimental data. We see that $\tchi_0$ exhibits a sharper feature at $\omega_\s\approx\omega_0$ as $T/T_\F$ decreases, a precursor of the singularity expected at $T=0$, which is characteristic of the edge of the particle-hole excitation continuum~\cite{NozieresPines1966}. Although there have been numerous indirect signatures of the Lindhard response, through inelastic electron scattering~\cite{schattschneider2012fundamentals}, X-ray scattering~\cite{comes1973evidence}, scanning tunneling microscopy~\cite{van1996direct}, and inelastic neutron scattering~\cite{renker1973observation}, this is, to the best of our knowledge, the first direct experimental observation of the Lindhard function.

\section{IV. Density response functions of the interacting Fermi gas}

Next, we study the density response of the interacting Fermi gas. 
In Fig.~\ref{FIG:2}, we show the measurements across a range of interaction strengths at fixed $T/\TF \approx 0.2$ (see Appendix A). As the interaction strength increases, the density response function becomes more symmetric and a sharper feature emerges, indicating the emergence of a collective mode in the interacting Fermi gas (for other recent examples of emergent hydrodynamics in Fermi systems see, \emph{e.g.},~\cite{brandstetter2023emergent,yan2024collective}). 

The first simple approach to describe the effect of interactions on the density response is to use the random phase approximation (RPA)~\cite{NozieresPines1966}: $\tchi_\mathrm{RPA}=\tchi_{\rm L}/(1-4\pi\kFa\tchi_{\rm L})$ for $|\kFa|\ll 1$. In Fig.~\ref{FIG:2}a, we show the measurements for two small and symmetric values of the interaction strength ($\kFa=\pm 0.12$). For a closer examination, we show in Fig.~\ref{FIG:2}b the differences between these response functions. The main effect of weak interactions within the RPA (for our typical regime of $T/T_\F$~\cite{zerosound}) is a shift of the spectrum that depends on the sign of $\as$; the spectrum remains asymmetric, without a well-defined resonance. The agreement with the RPA predictions (dashed lines) is excellent and shows that the mean-field interaction is the dominant contribution to the dynamical response for small $\kFa$.

In Fig.~\ref{FIG:2}c, we increase the interaction strength further~\cite{repul}. We see that the agreement with the RPA (dashed lines) quickly worsens. This is expected as the RPA only takes into account the mean-field interactions, while ignoring the effects of quasiparticle collisions, which play an increasingly important role as the interaction increases. We therefore adopt a theoretical description based on a linearized transport equation, which includes both the quasiparticle interaction and the quasiparticle collision effects~\cite{supp}: 
\begin{equation}\label{eq: transport equation}
    \left(\omega-\frac{\hbar\Bk\cdot\Bq}{m}\right) \delta\tilde{f}+\frac{\hbar\Bk\cdot\Bq}{m}f_\FD'(\epsilon_\Bk-\mu)(g\delta\tilde\rho+\tilde u_\mathrm{ext})=i\tilde{I}_{\text{coll}}^\text{(lin)},
\end{equation}
where $\delta\tilde{f}(\Bk,\Bq,\omega)$ is the Fourier transform of the deviation of the quasiparticle phase-space distribution $\delta\tilde{f}(\Bk,\Bq,\omega)\equiv \int\diff\Br\diff t~ \delta f(\Bk,\Br,t) e^{-i(\Bq\cdot\Br-\omega t)}$, where $\delta f(\Bk,\Br,t)=f(\Bk,\Br,t)-f_\FD(\epsilon_\Bk)$, $f_\FD(\epsilon)\equiv(1+\exp(\epsilon/(\kB T)))^{-1}$ is the Fermi-Dirac distribution, $\epsilon_\Bk$ is the energy of a quasiparticle of momentum $\hbar\Bk$, $g= 4\pi \hbar^2\as/m$ is the coupling strength, $\mu$ is the chemical potential, and $\tilde{I}_{\text{coll}}^\text{(lin)}$ is the linearized collision integral~\cite{supp}. In the language of Landau's theory of the FL, $g$ is related to the Landau parameters as $F_0^\s=-F_0^\a=2\kFa/\pi$ and $F_{l}=0$ for $l\geq1$ for small $\kFa$ (see Appendix C and~ \cite{lifshitz2013statistical}).

We identify in Eq.~\eqref{eq: transport equation} the three contributions to the response function. The first contribution [first term in Eq.~\eqref{eq: transport equation}] gives the Lindhard response (upon resummation). The second contribution is the mean-field interaction term (the first term in the second bracket of Eq.~\eqref{eq: transport equation}, $\propto g$); its magnitude is characterized by $\kFa$. Finally, the third is the collision integral (the right-hand side of Eq.~\eqref{eq: transport equation}, $\propto g^2$); it is characterized by a typical time scale $\tau_\text{coll}\equiv\hbar^3/(2m\as^2 \kB^2 T^2)$ and we define the \emph{collisional parameter} to be $\omega_0\tau_\text{coll}\equiv 2(\kFa)^{-2}(T/\TF)^{-2}(q_\text{ext}/\kF)$~\cite{supp}. There are two limiting regimes. In the collisionless regime $\omega_0\tau_\text{coll}\gg1$, the quasiparticles travel ballistically under the external and intrinsic conservative force $\nabla_\Br (u_\text{ext}(\Br,t)+g\rho(\Br,t))$ and sound can only propagate through the mean-field interactions. 
In the hydrodynamic regime, $\omega_0\tau_\text{coll}\ll1$, the role of collisions is crucial and Eq.~\eqref{eq: transport equation} reduces to the Navier–Stokes equations; ordinary (first) sound can propagate even for arbitrarily weak interactions provided that the sound wavelength or the temperature are large enough.

Note that Eq.~(\ref{eq: transport equation}) is a Vlasov-type equation for the quasiparticles since we take into account interactions only to first order; it also recovers the Boltzmann equation for particles at high $T$.  
At that mean-field level, the theory already includes the emergence of collective modes and nontrivial transport properties of the FL; if one wants to include renormalized properties of the quasiparticles, such as an effective mass, one would have to include next-order $(\kFa)^2$ corrections~\cite{lifshitz2013statistical}. The collision amplitude is calculated perturbatively \emph{ab initio} (without, \emph{e.g.}, a relaxation time approximation). In the thermodynamic limit, Eq.~\eqref{eq: transport equation} becomes a linear integral equation which we solve by expanding the quasiparticle distribution on a basis of orthogonal polynomials~\cite{supp}.

Despite the important role that the mean-field interactions play, we note that alone they do not capture the density response of an interacting Fermi system. Returning to Fig.~\ref{FIG:2}c-d, we alternatively show the calculated response functions only taking into account the effect of collisions (dotted lines) by solving Eq.~\eqref{eq: transport equation} with $g=0$: the agreement with the data is bad as well. It is only when we include both the mean-field and collision terms in solving Eq.~\eqref{eq: transport equation} (solid lines) that the agreement becomes very good. Remarkably, the data agree well with the theory up to $|\kFa|\approx0.66$ ($\omega_0\tau_\text{coll}\approx1$), without any adjustable parameters, showing a wide range of applicability of the first-order transport equation.

As the system enters the crossover regime,  $\omega_0\tau_\text{coll}\lesssim 1$, $\Re\tchi$ (resp. $\Im\tchi$) acquires a dispersive (resp. absorptive) shape, as expected from linearized hydrodynamics~\cite{NozieresPines1966,supp}.
Furthermore, the width decreases as $|\kFa|$ increases, \emph{i.e.}, the first-sound mode appears. Deviations between the measurements and the theory arise as the system passes through the crossover regime ($|\kFa|>0.66$, $\omega_0\tau_\text{coll}\lesssim1$). This deviation indicates the breakdown of Eq.~(\ref{eq: transport equation})~\cite{dynamicinst}. Extending Eq.~(\ref{eq: transport equation}) to $|\kFa| \gtrsim 1$, \emph{i.e.}, deriving a kinetic transport equation (or another first-principle description) for the quantum system in the strongly interacting regime is a major open challenge and the present data will provide a testing ground for such theories.

\section{V. Spectral properties and attenuation of sound}

To provide more insight into these results, we fit both measured and computed spectra with a generic form $\tchi_\text{fit}\propto 1/(\omega_\s-\omega_\p+i\Gamma/2)$, where $\omega_\p$ and $\Gamma$ are the peak frequency and width of the response, respectively (see~\cite{supp} for more details). These parameters provide a convenient effective characterization of the response functions. In Fig.~\ref{FIG:3} we show $\omega_\p$, $\Gamma$ and the quality factor $\mathcal{Q}\equiv\omega_\p/\Gamma$ as a function of $1/(\kFa)$, extracted from $\Re\tchi$ and $\Im\tchi$ as blue diamonds and red circles, respectively; the same quantities are similarly extracted from the theoretical spectra calculated from Eq.~\eqref{eq: transport equation} and shown as solid blue lines and red dotted lines, respectively. In the regime $\omega_0\tau_\text{coll}\gtrsim 1$ (where our theory Eq.~\eqref{eq: transport equation} works well), differences arise between the quantities extracted from $\Re\tchi$ and $\Im\tchi$ as the fitting form is not quantitatively accurate (especially as the response function is more asymmetric). In the hydrodynamic regime $\omega_0\tau_\text{coll}\lesssim 1$, they converge. 

We see that as $|\kFa|$ increases, $\omega_\p$ decreases, which is in line with the expectation that the compressibility increases monotonically with $|\kFa|$ in an attractive Fermi gas. Indeed, in the hydrodynamic limit, $\omega_\p$ is related to the sound velocity $c_1$ by $\omega_\p=c_1 q_\mathrm{ext}$; $c_1$ is linked to the isentropic compressibility $\kappa$ of the gas by the thermodynamic relation $c_1/v_\F=\sqrt{\kappa_0/(3\kappa)}$, where $\kappa_0$ is the compressibility of the ideal Fermi gas at $T=0$. At unitarity, our $c_1$ agrees with the measurement from~\cite{ku2012revealing,doi:10.1126/science.aaz5756} (the purple star).

  \begin{figure}[bt]
  \centering
  \includegraphics[width=1\columnwidth]{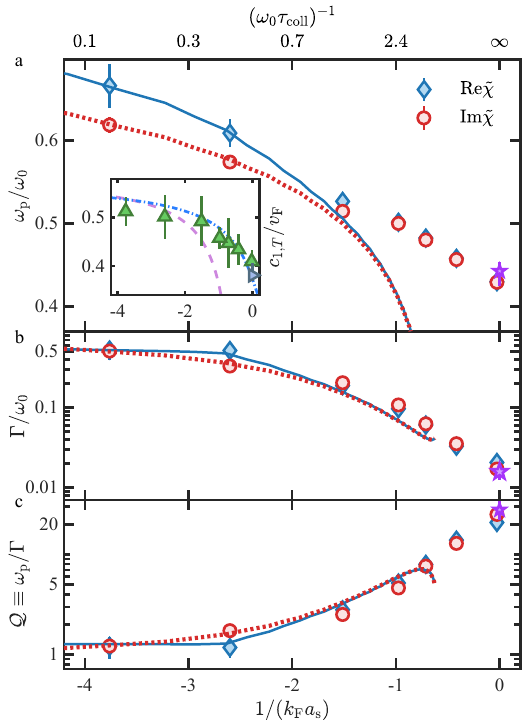}
  
  \caption{\textbf{Spectral properties extracted from the density response of a tunable Fermi liquid.} (\textbf{a}) Peak frequency $\omega_\mathrm{p}$, (\textbf{b}) width $\Gamma$, and (\textbf{c}) quality factor $\mathcal{Q}$ determined from response functions such as those shown in Fig.~\ref{FIG:2}. Circles and diamonds are data extracted from $\Im\tchi$ and $\Re\tchi$, respectively (see text). The blue solid and red dotted lines are from fits to the numerical solution of Eq.~\eqref{eq: transport equation}. The error bars are the uncertainties from the fit. The purple stars are the measurements from~\cite{doi:10.1126/science.aaz5756}. 
  The inset shows the isothermal speed of sound $c_{1,T}$ calculated from the isothermal compressibility: $c_{1,T}/\vF=\sqrt{\kappa_0/(3\kappa_T)}$, where $\kappa_T/\kappa_0\propto \tchi(\omega_\s/\omega_0\rightarrow0)$, see text. The pink dashed (resp. blue dot-dashed) line is calculated from the mean-field EoS (resp. EoS measured at low $T$~\cite{navon2010equation}). The gray triangle is the measurement from~\cite{ku2012revealing}.
  Note that by definition $\omega_0\tau_\text{coll}\equiv 2(\kFa)^{-2}(T/\TF)^{-2}(q_\text{ext}/\kF)$, where we use $q_\text{ext}/\kF\approx0.01$ and $T/\TF\approx0.2$.
  \label{FIG:3}}
\end{figure}

The low frequency limit of $\tchi$ allows us to extract the isothermal sound velocity, $c_{1,T}$ through the isothermal compressibility $\kappa_T$ since $\kappa_T/\kappa_0=-2\pi^2\tchi(\omega_\s\ll \omega_0)$~\cite{soundspeed}. In the inset of Fig.~\ref{FIG:3}a, we show the so-extracted $c_{1,T}$ as green triangles, together with the prediction of the low-$T$ equation of state (EoS)~\cite{navon2010equation} and the mean-field EoS (respectively the blue dot-dashed and pink dashed lines). Our measurement is consistent with the measurement from~\cite{ku2012revealing} (the gray triangle).

In Fig.~\ref{FIG:3}b, we plot $\Gamma$ versus $1/(\kFa)$. In the hydrodynamic regime, where $\Gamma$ captures the attenuation of sound~\cite{doi:10.1126/science.aaz5756,PhysRevA.81.053610}, the width decreases with increasing interaction strength, approaching the Heisenberg limit $\Gamma/\omega_0=\mathcal{O}(q_\text{ext}/\kF)$ in the strongly interacting limit~\cite{PhysRevLett.124.240403}. Note that our data at unitarity agree with a previous measurement derived from the sound diffusivity $D$, $\Gamma=D q_\text{ext}^2$~\cite{doi:10.1126/science.aaz5756}.
In the collisionless regime, the spectrum has a characteristic width $\Gamma\approx\omega_0$ corresponding to the width of the particle-hole continuum in the absence of zero sound~\cite{NozieresPines1966}. 
To assess the emergence of sound with increasing $\kFa$, we use $\mathcal{Q}$ as a measure of the collective mode's quality~\cite{PhysRevLett.124.150401}, which increases monotonically with $1/(\omega_0\tau_\text{coll})$ (see the top axis in Fig.~\ref{FIG:3}a), indicating the emergence of the first sound when one crosses from the collisionless to the hydrodynamic regime. 

\begin{figure*}[bt]
  \centering
  \includegraphics[width=1\textwidth]{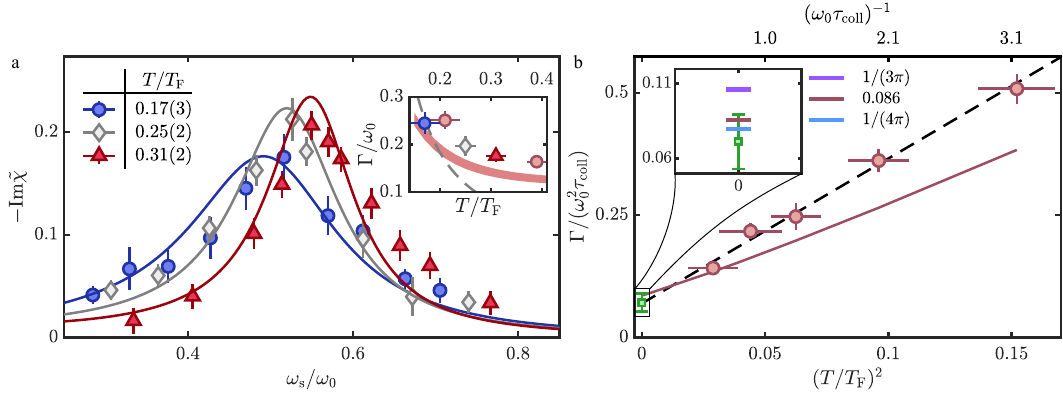}
  
  \caption{\textbf{Attenuation of the first sound in the hydrodynamic regime.} (\textbf{a}) Spectra of Im$\tchi$ at $\kFa\approx-0.67$ for various $T/T_\mathrm{F}$. The solid lines are the predictions from Eq.~\eqref{eq: transport equation}. Inset: $\Gamma/\omega_0$ versus $T/T_\F$. The gray dashed line is a reference of the damping of the first sound in a FL, $\Gamma/\omega_0\propto(T/\TF)^{-2}$. The red band is the prediction obtained from Lorentzian fits to the numerical solution of Eq.\eqref{eq: transport equation} including uncertainties on the gas parameters ($\approx3\%$ in $\kF$). (\textbf{b}) $\Gamma/(\omega_0^2\tau_\text{coll})$ versus $(T/\TF)^2$. The widths $\Gamma$ are extracted from  Lorentzian fits to the spectra shown in (a). The red line is the theoretical prediction and the black dashed line is a linear fit to the data (see text). The green square is the $T/\TF\rightarrow0$ limit from the fit, $a=0.07(2)_\text{stat}(3)_\text{sys}$ (see text). Inset: Zoomed-in version of (b) near $T/\TF=0$. The colored horizontal bars are theoretical predictions in the $T\rightarrow0$ limit: the Abrikosov-Khalatnikov calculation $1/(3\pi)$ (purple)~\cite{khalatnikov1958dispersion}, the prediction $1/(4\pi)$ (blue) from~\cite{bruun2005viscosity}, and the exact calculation $0.086$ (red)~\cite{exactres} based on~\cite{jensen1968exact,brooker1968transport}. 
  \label{FIG:4}}
\end{figure*}

We now turn to studying the attenuation of sound versus temperature. In Fig.~\ref{FIG:4}a, we show examples of $\Im\tchi$ measured at various $T/\TF$. In the inset, we show $\Gamma$ extracted from Lorentzian fits, as a function of $T$. We observe that $\Gamma/\omega_0$ decreases with increasing $T/\TF$. This behavior is qualitatively opposite to that of the strongly interacting Fermi gas~\cite{doi:10.1126/science.aaz5756,PhysRevLett.128.090402}, where sound attenuation, dominated by phonon-phonon collisions in the superfluid phase, increases monotonically with $T$, and also for the high-temperature weakly interacting Fermi gas, for which $\Gamma\propto \sqrt{T}$ (due to the two-body scattering)~\cite{khalatnikov1958dispersion,bruun2005viscosity}. It is also different from the sound attenuation in the FL hydrodynamic limit $\Gamma\propto 1/T^2$ (gray dashed line), which originates solely from the viscous damping~\cite{PhysRevLett.17.74,khalatnikov1958dispersion,NozieresPines1966}. This shows that this Fermi fluid is in a crossover regime from the FL to the high-temperature classical gas, where thermal conductivity also contributes to the damping of the sound. The results of the transport equation theory (colored lines and the red band) qualitatively agree with the experimental data, albeit with a small offset; this could be due to Eq.~\eqref{eq: transport equation} overestimating the quasiparticle collision integral or the quasiparticle interaction functions at $\kFa\approx-0.67$. 

For more insight, we show in Fig.~\ref{FIG:4}b the rescaled $\Gamma/(\omega_0^2\tau_\text{coll})$ ($\propto\Gamma T^2$), which is a universal function of $T/\TF$ at fixed $\kFa$ (see the solid red line) and is calculated from Eq.~\eqref{eq: transport equation} in the hydrodynamic limit~\cite{supp}. Specifically, $\Gamma/(\omega_0^2\tau_\text{coll})\approx a+b(T/\TF)^2$ at small $T/\TF$, where $a=\lim_{T/\TF\rightarrow0}2\hbar\eta/(3\EF\tau_\coll\rho_\eq)$ and $b$ can be calculated from the shear viscosity $\eta$ and the thermal conductivity~\cite{bulkvisc}. From the fit, we extrapolate to the $T/\TF\rightarrow 0$ limit and find the intercept to be $a=0.07(2)_\text{stat}(3)_\text{sys}$, where the statistical error bar is determined by bootstrapping and the systematic error takes into account uncertainties on the box volume and on the determination of $T$ from the EoS. In the inset of Fig.~\ref{FIG:4}b, we show that our measurement is in very good agreement with the exact result $0.086$ (red bar)~\cite{exactres,supp} and with an approximate analytical result $1/(4\pi)$ (blue bar)~\cite{bruun2005viscosity} but not with the Abrikosov-Khalatnikov result $1/(3\pi)$ (purple bar)~\cite{khalatnikov1958dispersion}. 
This quantitative agreement, especially on the non-trivial dependence of the width on temperature, highlights the applicability of the theory in studying this new crossover Fermi fluid.

\begin{figure*}[bt]
  \centering
  \includegraphics[width=1\textwidth]{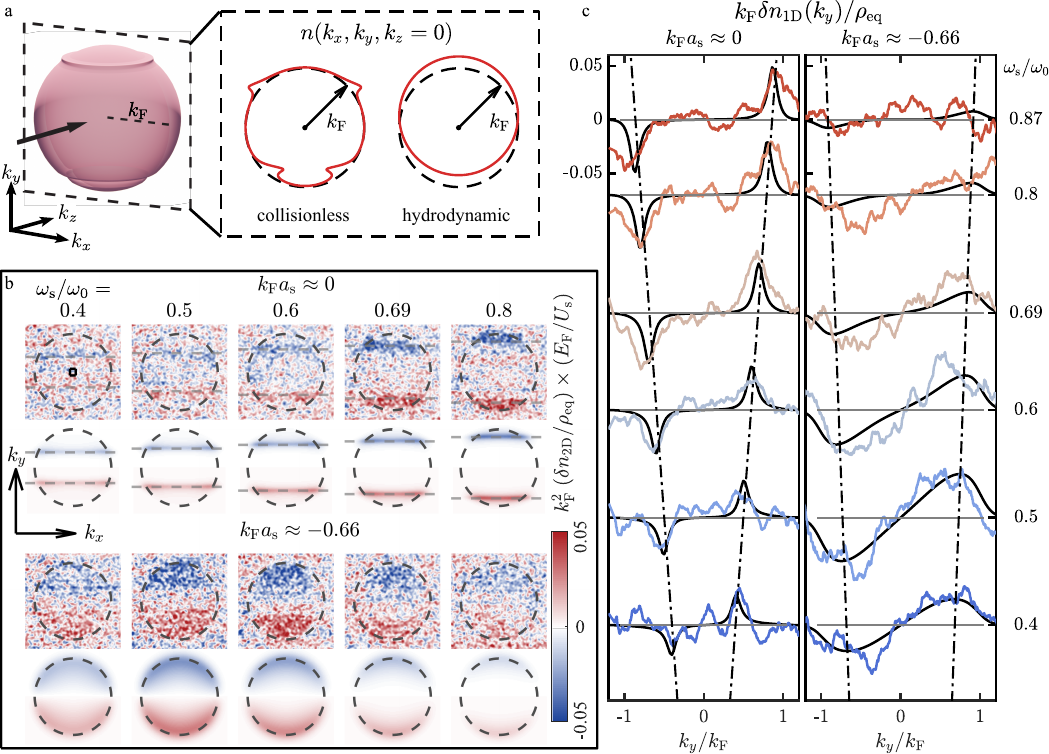}
  
  \caption{\textbf{Tomography of the quasiparticle excitations, from the collisionless to the hydrodynamic regime.} (\textbf{a}) Cartoon of the momentum-space excitation of the Fermi sea; the black arrow represents the imaging direction. The right panel shows the cut $n(k_x,k_y,k_z=0)$ of the momentum distribution for the excitations in the collisionless ($\omega_0\tau_\text{coll}\gg1$, left) and hydrodynamic ($\omega_0\tau_\text{coll}\ll1$, right) regimes. The black dashed circles of radius $\kF$ show the unperturbed Fermi sea. (\textbf{b}) The momentum excitation integrated along $k_z$: $\delta n_\mathrm{2D}(k_x,k_y)=\int\diff k_z~\delta n(\Bk)$, where $\delta n(\Bk)$ is the momentum distribution of the quasiparticles. The top panel and the bottom panel are galleries for the quasiparticle tomography for $\kFa\approx0$ and $\kFa\approx-0.66$, respectively. The top row in each panel is the experimental measurement for different $\ws/\omega_0$, and the bottom row is the theoretical predictions from FL theory Eq.~\eqref{eq: transport equation}. The color bar is the magnitude of $\kF^2 (\delta n_\mathrm{2D}/\rho_\eq)\times(\EF/\Us)$. The black rectangle in the first image shows the \emph{in situ} size of the box.  (\textbf{c}) The doubly integrated distribution $n_\mathrm{1D}(k_y)=\int\diff k_x~\delta n_{2\text{D}}$ for the two interaction strengths $\kFa\approx0$ (left) and $\kFa\approx-0.66$ (right) with various $\ws/\omega_0$ (see the right axis). The colored lines are the measurement and the black lines are the theoretical predictions. The black dash-dotted lines are the theoretical predictions for the peak positions.
  \label{FIG:5}}
\end{figure*}

\section{VI. Quasiparticle tomography}

Finally, observing directly the quasiparticle distribution (rather than averaged properties of the system such as current or magnetization) has long been sought~\cite{NozieresPines1966}. Here, we study this distribution as a function of $\kFa$. In Fig.~\ref{FIG:5}a we show a cartoon depiction of the quasiparticle distribution $n(\Bk)\propto\int\diff\Br~ f(\Bk,\Br)$. The Fermi sea (left) is perturbed along $y$, and imaged along $z$. We show on the right depictions of the central cuts $ n(k_x,k_y,k_z=0)$ expected in the collisionless and hydrodynamic limits. Note that the particle and quasiparticle momentum distributions coincide in the weakly interacting limit, up to corrections of order $(\kFa)^2$~ \cite{abrikosov2012methods}. Calculating the difference of momentum occupation is not trivial, especially at $T>0$; we estimate that at $T=0$, the depletion of particles within the Fermi sea is small, $\approx7\%$~\cite{lifshitz2013statistical,sartor1980self}; the finite temperature and line-of-sight integration will further smear out the particle momentum distribution and make the difference smaller~\cite{doggen2015momentum}.

Experimentally, we perform absorption imaging of the cloud after a long ToF, which gives access to the momentum distribution integrated along the line of sight $z$: $\delta n_{2\text{D}}(k_x,k_y)\propto\int\diff k_z~\delta n(\Bk)$, where $\delta n(\Bk)=n(\Bk)-n_\eq(\Bk)$ with $n_\eq(\Bk)=f_\FD(\epsilon_\Bk-\mu)$ is the equilibrium momentum distribution of the quasiparticles (see Appendix D). The \emph{in situ} size of the gas is small (marked by the black rectangle in the first image of Fig.~\ref{FIG:5}b) so that its effect on the measurement of the momentum distribution is negligible.
In Fig.~\ref{FIG:5}b we show a gallery of quasiparticle tomographies in the near-collisionless ($\kFa\approx0$ and $(\omega_0\tau_\text{coll})^{-1}\approx 0$, top) and near-hydrodynamic ($\kFa\approx-0.66$ and $\omega_0\tau_\text{coll}\approx3$, bottom) regimes. The shape of the quasiparticle excitations is qualitatively different in the two regimes. Interestingly, the excitation is more localized in momentum space in the collisionless regime than in the hydrodynamic one; while the opposite is true for the density response spectra.

The theoretical predictions are shown underneath the experimental measurements in Fig.~\ref{FIG:5}b~\cite{supp}. In the collisionless regime, excitations take the form $\delta n(\Bk) \propto \cos\theta/(\omega_\s/\omega_0-\cos\theta)$, where $\theta$ represents the polar angle with respect to the shaking direction. The pole at $\theta=\arccos(\omega_\s/\omega_0)$ reveals the nature of single particle-hole excitations. 
By contrast, in the hydrodynamic regime, $\delta n(\Bk) \propto \cos\theta$, preserving the isotropy of the background momentum distribution, showing that collisions tend to restore the equilibrium distribution faster than the drive period.

To facilitate a quantitative comparison, we turn to the 1D distributions $\delta n_{1\text{D}}(k_y)=\int\delta n_{2\text{D}}(k_x,k_y)\diff k_x$, shown in Fig.~\ref{FIG:5}c. The experimental data (colored lines) and the predictions (black solid lines) are in excellent agreement, without any free parameters. 
In the collisionless regime, the excitation is localized to a narrow peak in momentum space, whose position moves as $k_y/\kF=\omega_s/\omega_0$ (see black dot-dashed lines). Toward the hydrodynamic regime, the excitation is much broader and has a much weaker dependence on $\omega_\s/\omega_0$ (see black dot-dashed line on the rightmost plot). In the pure hydrodynamic regime, the excitations should have the same shape, with different amplitudes. Thus, the slight shifts of the peaks we observe here indicate that we are in the crossover regime~\cite{momentum}.

\section{VII. Conclusion and outlook}
We have shown that the tunable Fermi gas provides a versatile platform for studying Landau's theory of the Fermi liquid from first principles. Specifically, we measured the Lindhard function in a noninteracting Fermi gas and studied the emergence of the first sound in an FL by continuously tuning the interactions. We compared our measurements to a first-principle theory that incorporates both mean-field interactions and quasiparticle collisions and found good agreement from weak to moderate interaction strengths.

In the future, it would be interesting to attempt the observation of zero sound in the normal phase of the polarized Fermi gas~\cite{stringari2009density}. Furthermore, experiments on dipolar Fermi gases would provide a testbed for Landau's theory with long-range interactions~\cite{chan2010anisotropic}. Finally, theoretical refinements, particularly incorporating higher-order corrections into the Landau parameters~\cite{lifshitz2013statistical} and in the collision integrals, would not only allow one to take into account renormalization effects \emph{ab initio}, but also provide a controllable path toward understanding the transport properties (\emph{e.g.}, sound velocity, viscosity, and thermal conductivity) of the normal degenerate strongly interacting Fermi gas~\cite{PhysRevA.82.033619}. In particular, the measurement of $a\propto\Gamma T^2$ could provide a quantitative probe for the outstanding challenge of understanding the evolution of the Fermi liquid phase toward a non-Fermi liquid one in the strongly interacting regime of the BEC-BCS crossover~\cite{zwerger2011bcs,li2024observation}.

\section{Acknowlegment}
We thank L. Pricoupenko for discussions and S. Girvin, L. Glazman, F. Chevy and A. Tsidilkovski for comments on the manuscript. We also thank M. Zwierlein and P. Patel for sharing their data. This work was supported by the AFOSR (Grant No. FA9550-23-1-0605), NSF (Grant No. PHY-1945324), DARPA (Grant No. HR00112320038), and the David and Lucile Packard Foundation. H.K. acknowledges support from the EUR grant NanoX n°ANR-17-EURE-0009 in the framework of the “Programme des Investissements d’Avenir”. T.R. and H.K. thank Yale University for its hospitality.

\section{Appendix A: Preparation of the homogeneous Fermi fluid}
 We prepare a two-component Fermi gas of $^6$Li in a balanced mixture of the first and third lowest Zeeman sublevels, respectively labeled $\ket{\uparrow}=\ket{\frac{1}{2},+\frac{1}{2}}$ and $\ket{\downarrow}=\ket{\frac{3}{2},-\frac{3}{2}}$ in the low-field $\ket{F,m_F}$ basis ($F$ and $m_F$ are the total spin and its projection along the magnetic field axis). The interaction between the two components, characterized by the s-wave scattering length $\as$, is tuned by a broad Feshbach resonance at a bias magnetic field $B\approx 690$~G. The gas is initially trapped in a red-detuned crossed optical dipole trap (ODT). 
To prepare the weakly interacting (and the ideal) gas, the sample is first evaporatively cooled at a bias magnetic field $B\approx 284$~G, where $\as\approx -900 a_0$ ($a_0$ is the Bohr radius). The gas is then loaded in an optical cuboidal box shaped by digital micromirror devices using blue-detuned light of wavelength $\unit[639]{nm}$. The typical dimensions of the cuboid box are $L_x\times L \times L_z$ (where $L$ is along the drive direction $y$), with $L_x\approx\unit[114]{\mu m}$, $L\approx\unit[132]{\mu m}$ and $L_z\approx\unit[47]{\mu m}$. The magnetic field is then ramped to $B\approx 593$~G ($\as\approx 420 a_0$), where further evaporative cooling is performed in the box trap. At this stage, we typically have $N_{\uparrow} \approx N_{\downarrow} \approx 2.4 \times 10^5$ atoms per spin state at a temperature $T \approx 0.25T_\text{F}$ with $\TF \approx \unit[300]{nK}$. The magnetic field is then slowly ramped to the interaction field of interest. Similarly, to prepare a strongly interacting gas, we cool the atoms in the ODT and load them in the box at $B\approx 924$~G ($\as\approx-2700 a_0$), and subsequent evaporation is done at unitarity $B\approx 690$~G ($1/\as\approx0$). 

Thermometry is performed using two different methods depending on the interaction strength. For weak interactions ($|\kFa|\leq0.12$), we fit a Fermi-Dirac distribution to the density profile after a long time of flight (ToF). For stronger interactions, we measure the entropy $S$ at unitarity, before adiabatically ramping the magnetic field ramp to the field of interest; at unitarity, we use the calibrated radio-frequency shifts~\cite{mukherjee2019spectral} and speed of sound~\cite{doi:10.1126/science.aaz5756}. 
Since the entropy $S(T/\TF)$ for $\kFa<0$ is nearly indistinguishable from that of the ideal gas if $T>T_\text{c}$ (see,  \emph{e.g.},~\cite{haussmann2007thermodynamics,ku2012revealing}), we so extract $T/\TF$ at the field of interest (with an estimated systematic uncertainty of $\approx10\%$, due to the box volume). We verify the adiabaticity of the ramps by performing magnetic-field round trips and verifying that $S$ is unchanged.

\section{Appendix B: Measurement of the density response function}
We apply an oscillating force to the gas using a magnetic field gradient in the $y$ direction. We image the atoms \emph{in situ} along the imaging direction $z$; this provides the density response, $\rho_{\mathrm{2D}}(x,y,t)\equiv\int\diff z~\rho(\Br,t)$, which is both time- and spatially-resolved in the $x$ and $y$ directions. We then extract the center-of-mass (CoM) response of the gas along $y$, $\braket{y}(t)= \int \diff x\diff y\,(y \rho_{\mathrm{2D}})/\int\diff x\diff y\,\rho_\mathrm{2D}$. After a sufficiently long shaking time (typically $\unit[500]{ms}$), we find that the CoM response reaches a steady state, \emph{i.e.}, it oscillates at the same frequency as the drive $\ws$ and its oscillation amplitude does not depend on time. We typically measure the CoM at a sampling rate of $8\omega_\s$ and fit the CoM response to $A_y\sin(\ws t+\phi)+b$, where $A_y$, $\phi$, and $b$ are the amplitude, phase relative to the drive, and an offset of the response, respectively, from which we find the density response function $\tchi(\omega_\s)$ through Eq.~\eqref{eqn: tchi}.

\section{Appendix C: Landau parameters for a weakly interacting Fermi gas}
In Landau's theory of the Fermi liquid, the total energy of the system is postulated to be of the form~\cite{NozieresPines1966,baym2008landau,lifshitz2013statistical}
\begin{equation}
    \begin{aligned}
        \notag
        E&=E_0+\sum_{\Bk,\sigma=\{\uparrow,\downarrow\}}\epsilon_\sigma(\Bk)\delta n_\sigma(\Bk)\\
        &+\frac{1}{2}\sum_{\Bk,\Bk',\sigma,\sigma'}f_{\Bk\sigma,\Bk'\sigma'}\delta n_\sigma(\Bk)\delta n_{\sigma'}(\Bk'),
    \end{aligned}
\end{equation}
up to second order in the variation from equilibrium of the quasiparticle distribution $\delta n_\sigma(\Bk)$. The system is assumed to be spatially homogeneous so that $f_\sigma(\Br,\Bk)=n_\sigma(\Bk)$, where $n_\sigma(\Bk)$ is in general the spatially-averaged quasiparticle distribution; $\epsilon_\sigma(\Bk)$ is the equilibrium single-particle spectrum of the quasiparticle with spin index $\sigma$, $f_{\Bk\sigma,\Bk'\sigma'}$ are the quasiparticle interaction functions. For a weakly interacting Fermi gas, the total energy can be perturbatively calculated in a series of $\kFa$~\cite{lifshitz2013statistical}. Up to first order in $\kFa$, the total energy of the gas is
\begin{equation}
\begin{aligned}
    \notag
    E&=\sum_{\Bk,\sigma}\frac{\hbar^2 k^2}{2m}n_{\mathrm{eq},\sigma}(\Bk)+\frac{g}{\Vol}\sum_{\Bk,\Bk'}n_{\mathrm{eq},\uparrow}(\Bk)n_{\mathrm{eq},\downarrow}(\Bk'),
    \end{aligned}
\end{equation}
where $n_{\mathrm{eq},\sigma}$ is the Fermi-Dirac distribution. The function $f_{\Bk\sigma,\Bk'\sigma'}$ is then derived by a second variation of the total energy with the quasiparticle distribution:
\begin{align}
    f_{\Bk\uparrow,\Bk'\downarrow}=f_{\Bk\downarrow,\Bk'\uparrow}&=g/\Vol=f_{\Bk\Bk'}^\s-f_{\Bk\Bk'}^\a\notag,\\
    f_{\Bk\uparrow,\Bk'\uparrow}=f_{\Bk\downarrow,\Bk'\downarrow}&=0~~~~~=f_{\Bk\Bk'}^\s+f_{\Bk\Bk'}^\a\notag,
\end{align}
where we reexpressed them into $f_{\Bk\Bk'}^\s$ and $f_{\Bk\Bk'}^\a$, which can further be expanded in the basis of Legendre polynomials $f_{\Bk\Bk'}^{\s(\a)}=\sum_{l=0}^\infty f_l^{\s(\a)}P_l(\cos\theta)$ with $\theta$ is the angle between $\Bk$ and $\Bk'$. The conventional Landau parameters are defined by~\cite{NozieresPines1966,baym2008landau}
\begin{equation}\notag
    F_l^{\s(\a)}\equiv \frac{\Vol m^* \kF}{\pi^2\hbar^2}f_l^{\s(\a)},
\end{equation}
where $m^*$ is the effective mass. In our case of weakly interacting Fermi gas with first-order perturbation theory, the Landau parameters are given by
\begin{equation}\notag
    F_0^\s=-F_0^\a=\frac{2\kFa}{\pi},\quad F_l=0~\text{for }l\geq1,
\end{equation}
and the effective mass is the same as the particle mass $m^*=m$.

\section{Appendix D: Momentum-space tomography of the quasiparticle excitations}
We measure the momentum distribution in time of flight after an expansion in free space.
For the tomography measurement, we choose a drive time corresponding to $\omega_\s t=3.25$. We then turn off both the drive and the trap, let the gas expand for $t_\mathrm{TOF}=12$~ms and take an absorption image of the expanded cloud. The \emph{in situ} momentum distribution before the flight is then:
\begin{equation}\notag
    n_{\mathrm{2D}}(k_x,k_y)=\left(\frac{\hbar t_\mathrm{TOF}}{m}\right)^2 \rho ^\mathrm{TOF}_\mathrm{2D}\left(\frac{\hbar k_x}{m}t_\mathrm{TOF},\frac{\hbar k_y}{m}t_\mathrm{TOF}\right),
\end{equation}
where $\rho^\mathrm{TOF}_\mathrm{2D}(x,y)$ is spatial density of the expanded gas integrated along the line of sight $z$, and $\Bk=m\Br/(\hbar t_\mathrm{TOF})$  is the mapped wavevector. 

To avoid the convolution effect of the finite size of the box, we use for this measurement a small size box ($L\approx\unit[70]{\mu m}$) and a large density, so that $k_\mathrm{box}/\kF\approx0.15\ll1$. The black square in Fig.~\ref{FIG:5}b corresponds to the low momentum space cutoff $k_\mathrm{box}=mL/(\hbar t_\mathrm{TOF})$. Note that for our largest $|\kFa|$, the number of collision events (at $T/\TF\approx0.15$) during the expansion can loosely be estimated as $1/(\omega_0\tau_\text{coll})\approx 0.3$, so the expansion is only marginally ballistic.

To increase the signal quality of the subtracted distributions $n_\mathrm{2D}$, we extract the equilibrium (background) distribution from the data with the drive, by averaging the distributions over 4 phases of the drive ($0,~\pi/2,~\pi,~3\pi/2$). We apply a 2D Gaussian smoothing kernel with standard deviation $3$ image pixel ($\approx0.02\kF$) to the 2D distributions, and a Savitzky-Golay finite impulse response smoothing filter of polynomial order $1$ and frame length $31$ ($\approx 0.15\kF$) to the 1D distributions in Fig.~\ref{FIG:5}.

\endgroup

\newpage
\cleardoublepage
\setcounter{figure}{0}
\setcounter{equation}{0}
\newcounter{mycounter}
\setcounter{mycounter}{1}
\renewcommand{\thefigure}{S\arabic{figure}}
\newcommand{\RNum}[1]{\uppercase\expandafter{\romannumeral #1\relax.}}

\definecolor{blue(pigment)}{rgb}{0.2, 0.2, 0.6}
\definecolor{darkerblue}{rgb}{0.0, 0.0, 0.4}
\definecolor{darkblue}{rgb}{0.0,0.0,0.5}
\definecolor{darkgreen}{rgb}{0.0,0.4,0.0}
\hypersetup{
   colorlinks,
   linkcolor=blue(pigment),
    citecolor=darkgreen,
   urlcolor=darkblue
}

\renewcommand{\theequation}{S\arabic{equation}}
\setcounter{secnumdepth}{3}

\onecolumngrid
{\centering\bf  \large Supplementary material for:\\
\vspace{0.1cm}
\centering
``Emergence of Sound in a Tunable Fermi Fluid''\\}
\onecolumngrid
\tableofcontents
\vspace{1em}

\section{The weakly interacting Fermi gas as a Fermi liquid}\label{Fermi}
\stepcounter{mycounter}

\subsection{Basic formalism of the Fermi liquid}
Landau's theory postulates the existence of quasiparticles whose quantum number (\emph{e.g.} the momentum $\Bp=\hbar\Bk$ in a homogeneous system) is the same as for the particles in the noninteracting system. This originates from the assumption of adiabatic turning on of the interactions. The state of the system is then defined by the distribution of quasiparticles $n_\sigma(\Bp)$ of spin $\sigma$, which deviates from the equilibrium distribution $n_{\eq,\sigma}(\Bp)$. The quasiparticle energy at equilibrium $\epsilon_{\rm eq,\sigma}(\Bp)$ (\emph{i.e.}, the energy cost of adding a quasiparticle of spin $\sigma$ in the equilibrium state) and their interaction function $f_{\Bp\sigma,\Bp'\sigma'}$ are defined from the expansion of the total energy $E$ in the vicinity of equilibrium:
\be
E=E_0+\sum_{\Bp,\sigma} \epsilon_{\rm eq,\sigma}(\Bp) \delta n_\sigma(\Bp)+\frac{1}{2}\sum_{\Bp\sigma,\Bp'\sigma'} f_{\Bp\sigma,\Bp'\sigma'} \delta n_{\sigma}(\Bp) \delta n_{\sigma'}(\Bp'),
\ee
where $E_0$ is the ground state energy and $\delta n_\sigma(\Bp)\equiv n_\sigma(\Bp)-n_{\eq,\sigma}(\Bp)$. More explicitly:
\bea
\epsilon_{\rm eq,\sigma}(\Bp) &=& \left.\frac{\partial E}{\partial n_\sigma(\Bp)}\right\vert_{ n=n_\eq} \label{epsilon0}  \\
f_{\Bp\sigma,\Bp'\sigma'}  &=& \left.\frac{\partial^2 E}{\partial n_\sigma(\Bp)\partial n_{\sigma'}(\Bp')}\right\vert_{ n=n_\eq}. \label{fpp}
\eea
In Landau's theory of the FL, quasiparticles are well defined only in the vicinity of the Fermi surface, and their equilibrium distribution is a Fermi-Dirac distribution at temperature $T$ and chemical potential $\mu_\sigma$:
\be
n_{\rm eq,\sigma}(\Bp)=\frac{1}{1+\eee^{(\epsilon_{\rm eq,\sigma}(\Bp)-\mu_\sigma)/(\kB T)}},
\ee
where their dispersion relation can be expanded in an isotropic system as
\be
\epsilon_{\rm eq,\sigma}(\Bp)-\mu_\sigma=\frac{p_{\F,\sigma}}{m^*_\sigma}(p-p_{\F,\sigma}).
\ee
Here $p_{\F,\sigma}=\hbar(6\pi^2\rho_{\eq,\sigma})^{1/3}$ is the Fermi momentum, $\rho_{\eq,\sigma}$ the equilibrium density of the spin $\sigma$ component, $m^*_\sigma$ is the effective mass.

\subsection{Fermi liquid parameters of the weakly interacting Fermi gas}
We consider a homogeneous gas of spin-1/2 fermions interacting via s-wave contact interactions and derive the quasiparticle dispersion and interaction function from first principles. The Hamiltonian of the system is
\be\label{Hamiltonian}
\hat H_\s=\sum_{\Bp,\sigma}\epsilon_p\hat c_{\Bp,\sigma}^\dagger \hat c_{\Bp,\sigma}+ \frac{g}{\Vol}\sum_{\Bp_1,\Bp_2,\Bp_3,\Bp_4} \delta_{\Bp_1+\Bp_2,\Bp_3+\Bp_4}\hat c_{\Bp_1,\uparrow}^\dagger \hat c_{\Bp_2,\downarrow}^\dagger \hat c_{\Bp_3,\downarrow} \hat c_{\Bp_4,\uparrow},
\ee
where $\sigma=\{\uparrow,\downarrow\}$, $\epsilon_p=p^2/(2m)$ is the kinetic energy, $m$ is the mass of the particle, $g$ is the coupling constant, $\Vol$ is the volume of the system, and $\hat c_{\Bp,\sigma}$ (resp. $\hat c_{\Bp,\sigma}^\dagger$) annihilates (resp. creates) a fermion of momentum $\Bp$ and spin $\sigma$.
To first order in perturbation theory, we relate the coupling constant $g$ to the s-wave scattering length in the Born approximation $g=4\pi\hbar^2 \as/m$, and the energy levels of the interacting Hamiltonian are then~\cite{lifshitz2013statisticalSM}
\be
E_1=\meanv{\hat H}=\sum_{\Bp,\sigma} \frac{p^2}{2m} n_{\sigma}(\Bp)+\frac{g}{\Vol}\sum_{\Bp_1,\Bp_2} n_{\uparrow}(\Bp_1)  n_{\downarrow}(\Bp_2). \label{E1eq}
\ee
The average is taken over the eigenstate of the noninteracting Fermi gas, which can be described by the momentum distribution $n_{\sigma}(\Bp)=\meanv{\hat c_{\Bp,\sigma}^\dagger \hat c_{\Bp,\sigma}}$.
This expression shows that to first order, the quasiparticles have the same
distributions as the original particles, while their equilibrium dispersion relation is modified by a mean-field energy
\be
\epsilon_{{\rm eq},\sigma}(\Bp)=\frac{p^2}{2m}+g\rho_{{\rm eq},-\sigma}\label{EFermigas},
\ee
where $-\sigma$ is the spin opposite to $\sigma$, and the quasiparticle interaction function is $
f_{\Bp\uparrow,\Bp'\downarrow}=g/\Vol$.
The Landau parameters, \emph{i.e.} the Legendre components
of the interaction function, are then restricted to $l=0$: $F_0^{\rm s}=-F_0^{\rm a}= 2\kFa/\pi$, and $F_{l}=0$ for $l\geq1$.
These simple Fermi liquid parameters allow for an exact solution of the transport equation, see section~\ref{sec: hydrodynamic}.

\subsection{Equilibrium properties at finite temperature}

The distribution of the equilibrium state for the equal spin-populations density system is
\be
n_{\rm eq}(\Bp)=\frac{1}{1+\eee^{({p^2}/{2m}+g\rho_{{\rm eq}}-\mu)/(\kB T)}},
\ee
where $\mu_\uparrow=\mu_\downarrow=\mu$ and $\rho_{{\rm eq},\uparrow}=\rho_{\rm eq,\downarrow}=\rho_{\rm eq}$. In the following sections, we will drop the spin index for simplicity.
The equation of state relates $\mu$ to the equilibrium density $\rho_{\rm eq}$:
\be
\rho_{\rm eq}=\frac{1}{\Vol}\sum_\Bp n_{\rm eq}(\Bp),\label{eqdetat}
\ee
from which we get
\be
\mu=\mu_0(T)+g\rho_{{\rm eq}},
\ee
where $\mu_0(T)$ is the chemical potential of the ideal gas.
The quasiparticle distribution is thus the same as that of the noninteracting Fermi since
\be
n_{\rm eq}(\Bp)=\frac{1}{1+\eee^{({p^2}/{2m}-\mu_0)/(\kB T)}}.
\ee
Thus, the first corrections to the distribution function appear at order $(\kFa)^2$~\cite{abrikosov2012methodsSM}.
The associated fugacity
$z_0=\eee^{\beta\mu_0}$ is implicitly related to $T/T_F$ by
\be
{\frac{T}{T_F}}=\bbcro{\frac{3\sqrt{\pi}}{4}\mathcal{F}_{1/2}(\beta\mu_0)}^{-2/3},
\ee
where $\mathcal{F}_{n}(m)=\frac{1}{\Gamma(n+1)}\int_0^{+\infty}\frac{t^n \dd t}{\eee^{t-m}+1}$ is the Fermi-Dirac integral. 

The equation of state Eq.~\eqref{eqdetat} yields the isothermal compressibility $\kappa_T\equiv\bbcro{\rho_\eq^2(\partial\mu/\partial\rho_\eq)_{T}}^{-1}$:
\be
\frac{ \kappa_T}{\kappa_0(0)}=\frac{1}{\frac{\kappa_0(0)}{\kappa_0(T)}+\frac{2}{\pi}\kFa}
\label{kappaT}
\ee
where the isothermal compressibility $\kappa_0(T)$ of the ideal gas at temperature $T$ is
\be
\kappa_0(T)=-\bb{\frac{\pi}{6}}^{1/3}\frac{\text{Li}_{1/2}(-z_0)}{(-\text{Li}_{3/2}(-z_0))^{1/3}}\kappa_0(0),
\ee
$\kappa_0(0)=3/(2\rho_\eq \EF)$, and $\text{Li}_{n}$ is the polylogarithm of order $n$, defined as $\text{Li}_{n}(-z)=-\mathcal{F}_{n-1}(\ln z)$. 

\section{Density-density linear response theory}
\stepcounter{mycounter}

\subsection{Linear response theory}
Suppose that we apply a periodic density perturbation to the system so that the total Hamiltonian reads
\begin{equation}
    \hat{H}=\hat{H}_\s+\hat{V}(t),
\end{equation}
where $\hat{H}_\s$ is the Hamiltonian in Eq.~\eqref{Hamiltonian} of the system without the drive. The perturbation $\hat{V}(t)$ is
\begin{equation}
    \begin{aligned}
        \hat{V}(t)&=\int\diff\Br~u_\text{ext}(\Br,t)\hat{\psi}^\dagger(\Br)\hat{\psi}(\Br),
    \end{aligned}
\end{equation}
where $u_\ext(\Br,t)=-(\Us/L)y\sin(\ws t)$ is the external driving potential, $L$ the length of the box along the direction $y$ of the drive; $\Us$ and $\ws$ are respectively the amplitude and frequency of the drive. The operators $\hat{\psi}^\dagger(\Br)$ and $\hat{\psi}(\Br)$ respectively create and annihilate a fermion of either spin at position $\Br$. Using Kubo's formula~\cite{kubo1957statistical}, the response of the density $\hat{\rho}(\Br)\equiv\hat{\psi}^\dagger(\Br)\hat{\psi}(\Br)$ to the perturbation is
\begin{equation}\label{eqn: LR real space}
    \delta\rho(\Br,t)\equiv \braket{\hat{\rho}}-\braket{\hat{\rho}}_\text{eq}=\int\diff\Br\int\diff t'~\chi^\R_{\rho\rho}(\Br-\Br',t-t')u_\text{ext}(\Br',t').
\end{equation}
Here $\chi_{\rho\rho}^\R(\Br,t)$ is the retarded density-density response function, defined as
\begin{equation}
    \chi^\R_{\rho\rho}(\Br-\Br',t-t')=-\frac{i}{\hbar}\Theta(t) \braket{[\hat{\rho}(\mathbf{r},t), \hat{\rho}(\Br',t')]}_\text{eq},
\end{equation}
where $\Theta(t)$ is the Heaviside function, the operators are evaluated in the interaction picture, and $\braket{\cdot}_\text{eq}$ is the ensemble average over the equilibrium distribution (without the perturbation). After a Fourier transform, Eq.~\eqref{eqn: LR real space} becomes
\begin{equation}\label{eqn: LR Fourier space}
    \delta\tilde{\rho}(\Bq,\omega)=\tchi_{\rho\rho}(\Bq,\omega)\tilde{u}_\text{ext}(\Bq,\omega),
\end{equation}
where $\tilde{X}$ denotes the Fourier transform of quantity $X$.

\subsection{Center-of-mass response under periodic drive}

Using our box boundary conditions, we decompose the driving potential as 
\begin{equation}\label{eqn: shaking first mode}
    \begin{aligned}
        u_\text{ext}(\Br,t)&=-\Us\sum_{n>0,\text{ odd}}(-1)^{\frac{n-1}{2}}\frac{4}{(n\pi)^2}\sin\left(\frac{n\pi y}{L}\right)\sin(\ws t)\\
        &\approx-\frac{4\Us}{\pi^2}\sin\left(\frac{\pi y}{L}\right)\sin(\ws t).
    \end{aligned}
\end{equation}
In the second line, we only consider the lowest-lying mode, $n=1$. After a Fourier transform, Eq.~\eqref{eqn: shaking first mode} becomes
\begin{equation}\label{eqn: uext Fourier}
    \tilde{u}_\text{ext}(\Bq,\omega)=\frac{\Us}{\pi^2}(2\pi)^4\left[\delta(\Bq-q_\ext\hat{y})\delta(\omega+\ws)+
    \delta(\Bq+q_\ext\hat{y})\delta(\omega-\ws)-
    \delta(\Bq-q_\ext\hat{y})\delta(\omega-\ws)-
    \delta(\Bq+q_\ext\hat{y})\delta(\omega+\ws)\right],
\end{equation}
where $q_\ext=\pi/L$ and $\hat{y}$ is the unit vector along the direction $y$. From Eq.~\eqref{eqn: LR Fourier space}, we define the dimensionless density response function at the wavevector $\mathbf{q}=q_\text{ext}\hat{y}$ as 
\be
\tchi(\omega)\equiv\frac{2\EF}{\kF^3}\tchi_{\rho\rho}(q_\ext\hat{y},\omega),
\ee
where $\kF=p_\F/m$ is the Fermi wavevector, and $\EF=\hbar^2\kF^2/(2m)$ is the Fermi energy.
By plugging Eq.~\eqref{eqn: uext Fourier} into Eq.~\eqref{eqn: LR Fourier space} and applying an inverse Fourier transform, we derive the density response in real space: 
\begin{equation}
    \delta\rho(\Br,t)=-12\rho_\eq \frac{\Us}{\EF}\left(\Re\tchi(\ws) \sin(\ws t) - \Im\tchi(\ws) \cos(\ws t)\right)\sin\left(\frac{\pi y}{L}\right),
\end{equation}
where we used that $\tchi_{\rho\rho}(\Bq,\omega)=\tchi_{\rho\rho}(-\Bq,\omega)=[\tchi_{\rho\rho}(\Bq,-\omega)]^*$~\cite{NozieresPines1966SM}.

In the experiment, we measure the center of mass (CoM) $\braket{y}$ along the shaking direction, which is given by 
\begin{equation}\label{CoMy LR}
    \begin{aligned}
        \braket{y}(t)&\equiv\frac{\int\diff \Br ~y\delta\rho(\Br,t)}{\int\diff \Br ~\rho(\Br,t)}\\
        &=-L\frac{24}{\pi^2}\frac{\Us}{\EF}\left(\Re\tchi \sin(\ws t)-\Im\tchi\cos(\ws t)\right)=A_y\sin(\ws t+\phi),
    \end{aligned}
\end{equation}
where the CoM oscillation amplitude $A_y$ and phase $\phi$ are
\begin{align}
    \frac{A_y}{L}&=\frac{24}{\pi^2}\frac{\Us}{\EF}\left|\tchi(\ws)\right|,\label{eqn: Ay}\\
    \tan\phi&=-\frac{\Im\tchi(\ws)}{\Re\tchi(\ws)}\label{eqn: tanphi}.
\end{align}

\section{Transport equation}
\stepcounter{mycounter}
In this section, we introduce the transport equation for the quasiparticles, which we use to calculate the density response functions.
\subsection{Local balance of the quasiparticle distribution}
We describe the system \textit{near equilibrium} with a \textit{local} quasiparticle distribution $f(\Bp,\rr,t)=f_\uparrow(\Bp,\rr,t)=f_\downarrow(\Bp,\rr,t)$ and a local energy $\epsilon(\Bp,\rr,t)$.
This is justified because in practice we work in the weak-perturbation ($\Us/\EF\ll 1$) and long-wavelength ($q_\ext\ll\kF$ - typically we have $\kF L\gtrsim 250$ in the experiment) limits and we only consider spin-independent perturbations. In these limits, the typical length scale and frequency scale for the density response is $1/q_\ext$ and $\omega_0=q_\ext\vF$, respectively, and we can write the quasiparticle energy as~\cite{baym2008landauSM}:
\bea
\epsilon(\Bp,\rr,t)&=&\epsilon_{\rm eq}(\Bp)+g\delta \rho(\rr,t)\\
&=&\frac{p^2}{2m}+g\rho(\Br,t)\label{E1},
\eea
where we have used Eq.~\eqref{EFermigas}, with $\delta\rho(\Br,t)=(1/\Vol)\sum_\Bp\delta f(\Bp,\Br,t)$ and $\delta f(\Bp,\Br,t)=f(\Bp,\Br,t)-n_\eq(\Bp)$. The dynamics is determined by a kinetic equation, which is derived by balancing the quasiparticle fluxes in a phase-space volume $\dd^3 r\dd^3 p$: quasiparticles travel with a velocity ${\partial \epsilon}/{\partial \Bp}$
under the influence of the intrinsic force $-{\partial \epsilon}/{\partial \rr}$ and the drive force
$-{\partial u_{\ext}}/{\partial \rr}$: 
\be
\frac{\partial f}{\partial t}  +\frac{\partial \epsilon}{\partial \Bp}  \frac{\partial f}{\partial \rr} - \frac{\partial (\epsilon+u_{\ext})}{\partial \rr} \frac{\partial f}{\partial \Bp} =I_{\rm coll}.
\label{Boltzmann1}
\ee
Using Eq.~\eqref{E1}, this equation becomes
\be
\frac{\partial f}{\partial t}  +\frac{\Bp}{m}  \frac{\partial f}{\partial \rr} - \bb{g \frac{\partial \rho}{\partial \rr} +\frac{\partial u_{\ext}}{\partial \rr}}  \frac{\partial f}{\partial \Bp} =I_{\rm coll}. \label{boltzamannweak}
\ee
The collision integral $I_{\rm coll}(\Bp,\Br,t)$ counts the number of quasiparticles with momentum $\Bp$ that are annihilated or created at position $\rr$ and time $t$:
\begin{multline}
I_{\rm coll}(\Bp,\Br,t)=2\pi\sum_{\Bp_2,\Bp_3,\Bp_4} W(\Bp;\Bp_2\vert\Bp_3;\Bp_4)\delta_{\Bp+\Bp_2,\Bp_3+\Bp_4}\delta(\epsilon({\Bp,\rr,t})+\epsilon\bb{\Bp_2,\rr,t}-\epsilon\bb{\Bp_3,\rr,t}-\epsilon\bb{\Bp_4,\rr,t}) \\
\bbcro{(1-f(\Bp,\Br,t))(1-f(\Bp_2,\rr,t))f(\Bp_3,\rr,t)f(\Bp_4,\rr,t)-f(\Bp,\Br,t)f(\Bp_2,\rr,t)(1-f(\Bp_3,\rr,t))(1-f(\Bp_4,\rr,t))}
\end{multline}
where $W$ is the probability of the collision $(\Bp,\sigma;\Bp_2,-\sigma)\to(\Bp_3,\sigma;\Bp_4,-\sigma)$; we assumed only elastic collisions. The coupling amplitude associated with $W$ is in general related to $f_{\Bp\sigma,\Bp'\sigma'}$ by an integral equation (see Eq.~(1.112) in \cite{NozieresPines1966SM}). In our weakly interacting gas, it is simply $W=g^2/\Vol^2$.

\subsection{Weak-drive limit: linearized transport equation}

We linearize Eq.~\eqref{boltzamannweak} around the equilibrium distribution $n_{\eq}(\Bp)$ assuming $\delta f(\Bp,\Br,t)$ is small and proportional to $u_\ext$. 
Up to order $\mathcal{O}(u_\ext)$, we get
\be
\frac{\partial \delta f}{\partial t}  +\frac{ \Bp}{m}  \frac{\partial \delta f}{\partial \rr} - \bb{g \frac{\partial \delta\rho}{\partial \rr} +\frac{\partial u_\ext}{\partial \rr}} \frac{\partial n_{\rm eq}}{\partial \Bp} =I_{\rm coll}^{\rm (lin)}.
\label{Boltzmann2}
\ee
Note that ${\partial n_{\rm eq}}/{\partial \Bp}=f_\text{FD}'(\epsilon_{\rm eq}(\Bp)-\mu)\times(\Bp/m)=f_\text{FD}'(\epsilon_p-\mu_0(T))\times(\Bp/m)$ can be expressed in terms of the Fermi-Dirac distribution $f_\text{FD}(\epsilon)=1/(1+\eee^{\epsilon/(\kB T)})$ (and $f_\text{FD}'(\epsilon)\equiv \diff f_\text{FD}/\diff \epsilon$). 
The linearized collision integral $I_{\rm coll}^{\rm (lin)}$ takes the form
\be
I_{\rm coll}^{\rm (lin)}=-\bbcro{\Gamma(\Bp)\delta f(\Bp,\Br,t)+\frac{1}{\Vol}\sum_{\Bp'} (E(\Bp',\Bp)-2S(\Bp',\Bp))\delta f(\Bp',\rr,t)},\label{Icolllin}
\ee
where $\Gamma(\Bp)$ is the quasiparticle scattering rate:
\be
\Gamma(\Bp)=2\pi\frac{g^2}{\Vol^2} \sum_{\Bp_2,\Bp_3,\Bp_4} \delta_{\Bp+\Bp_2,\Bp_3+\Bp_4} 
\delta\bb{\frac{p^2}{2m}+\frac{p_2^2}{2m}-\frac{p_3^2}{2m}-\frac{p_4^2}{2m}}  \bbcro{\vphantom{\frac{1}{2}}n_{\rm eq}(\Bp_2)(1-n_{\rm eq}(\Bp_3)-n_{\rm eq}(\Bp_4))+n_{\rm eq}(\Bp_3) n_{\rm eq}(\Bp_4)}.
\label{Gammap}
\ee
The two collision kernels $E(\Bp,\Bp')$ and $S(\Bp,\Bp')$ are defined as
\begin{multline}
E(\Bp,\Bp')=2\pi\frac{g^2}{\Vol} \sum_{\Bp_3,\Bp_4}\delta_{\Bp+\Bp',\Bp_3+\Bp_4}\delta\bb{\frac{p^2}{2m}+\frac{p'^2}{2m}-\frac{p_3^2}{2m}-\frac{p_4^2}{2m}}\\
\times\bbcro{\vphantom{\frac{1}{2}}n_{\rm eq}(\Bp')(1-n_{\rm eq}(\Bp_3)-n_{\rm eq}(\Bp_4))+n_{\rm eq}(\Bp_3) n_{\rm eq}(\Bp_4)} \label{Epp}
\end{multline}
and
\begin{multline}
S(\Bp,\Bp')=2\pi\frac{g^2}{\Vol} \sum_{\Bp_2,\Bp_4}\delta_{\Bp+\Bp_2,\Bp'+\Bp_4}\delta\bb{\frac{p^2}{2m}+\frac{p_2^2}{2m}-\frac{p'^2}{2m}-\frac{p_4^2}{2m}}\\
\times\bbcro{\vphantom{\frac{1}{2}}n_{\rm eq}(\Bp_2)(1-n_{\rm eq}(\Bp')-n_{\rm eq}(\Bp_4))+n_{\rm eq}(\Bp') n_{\rm eq}(\Bp_4)}, \label{Spp}
\end{multline}
where for $E(\Bp,\Bp')$ the pair of momenta $(\Bp,\Bp')$ are on the same side of the collision (\emph{i.e.}, both are the initial states or the final states of the collision event), and for $S(\Bp,\Bp')$ they are on opposite sides.

We now write Eq.~\eqref{Boltzmann2} in Fourier space. The drive is $u_\ext(\rr,t)= (1/\Vol) \sum_\qq \int\diff\omega/(2\pi)~\tilde{u}_\ext(\qq,\omega) \eee^{\ii(\qq\cdot\rr-\omega t)}$, the quasiparticle distribution is $\delta f(\Bp,\Br,t)=(1/\Vol)\sum_\qq \int\diff\omega/(2\pi)~\eee^{\ii(\qq\cdot\rr-\omega t)} \delta\tilde{f}(\Bp,\qq,\omega)$, the density distribution is $\delta\rho(\Br,t)=(1/\Vol)\sum_\qq \int\diff\omega/(2\pi)~\eee^{\ii(\qq\cdot\rr-\omega t)} \delta\tilde{\rho}(\qq,\omega)$, and the collision integral is $I_{\rm coll}^{\rm (lin)}(\Bp,\Br,t)=(1/\Vol)\sum_\qq \int\diff\omega/(2\pi)~\eee^{\ii(\qq\cdot\rr-\omega t)} \tilde{I}_{\rm coll}^{\rm (lin)}(\Bp,\qq,\omega)$. This yields:
\be
\bb{\omega-\frac{\Bp\cdot\qq}{m}}  \delta\tilde{f}({\Bp,\Bq,\omega}) +\frac{\Bp\cdot\qq}{m}f_\FD'(\epsilon_\eq(\Bp)-\mu) \bb{g \delta\tilde{\rho}(\qq,\omega)+\tilde{u}_\ext(\qq,\omega)}=\ii \tilde{I}_{\rm coll}^{\rm (lin)}(\Bp,\qq,\omega).
\label{BoltzmannFourier}
\ee

In practice, we restrict those decompositions to the two opposite Fourier components $\Bq=\pm q_\ext\hat{y}$ and $\omega=\pm\ws$ [see Eq.~\eqref{eqn: uext Fourier}]. 
The density response function, defined by Eq.~\eqref{eqn: LR Fourier space}, is
\begin{equation}\label{chiBoltz}
    \tilde{\chi}_{\rho\rho}(\Bq,\omega)=\frac{\delta\tilde{\rho}(\Bq,\omega)}{\tilde{u}_\ext(\Bq,\omega)}=\frac{1}{\Vol}\sum_\Bp \frac{\delta \tilde{f}(\Bp,\Bq,\omega)}{\tilde{u}_\ext(\Bq,\omega)}.
\end{equation}

\subsection{Change of variable}
At $T\ll\TF$, $\delta\tilde{f}$ is usually sharply peaked at the Fermi level $|\Bp|=p_\F$. We perform a change of variables from $\delta \tilde{f}$ to $\tilde{\nu}$, where the new variable $\tilde{\nu}$ is defined by
\be
\delta\tilde{f}(\Bp,\Bq,\omega)=f_\FD'(\epsilon_\eq(\Bp)-\mu)\tilde{\nu}(\Bp,\Bq,\omega) \tilde{u}_\ext(\Bq,\omega),
\label{chgvar}
\ee
in order to work with a smooth function across the Fermi level. The linearized transport equation Eq.~\eqref{BoltzmannFourier} becomes
\be
\bb{\omega-\frac{\Bp\cdot\qq}{m}} \tilde{\nu}({\Bp})+\frac{g}{(2\pi\hbar)^3}  \frac{\Bp\cdot\qq}{m}\int\dd^3 p' f_\FD'(\epsilon_\eq(\Bp')-\mu)\tilde{\nu}({ \Bp'})+\ii \int{\dd^3 p'}{ \mathcal{N}_{\rm coll}({\Bp},{\Bp'})}\tilde{\nu}({\Bp'})=- \frac{\Bp\cdot\qq}{m},
\label{limitethermo}
\ee
where
\be
\mathcal{N}_{\rm coll}({\Bp},{\Bp'})=\Gamma(\Bp)\delta(\Bp-\Bp')+\frac{E(\Bp,\Bp')-2S(\Bp,\Bp')}{(2\pi\hbar)^3} \label{Ncoll}
\ee
is the kernel of the linearized collision integral. Note that since
\begin{equation}
    E(\Bp,\Bp') = \frac{f_\FD'(\epsilon_{\eq}(\Bp')-\mu)}{f_\FD'(\epsilon_{\eq}(\Bp)-\mu)}E(\Bp',\Bp)
\end{equation}
[and the same holds for $S(\Bp,\Bp')$], the momenta $\Bp$ and $\Bp'$ end up swapped from Eq.~\eqref{Icolllin} to Eq.~\eqref{Ncoll}. 
Finally, the density response function Eq.~\eqref{chiBoltz} becomes
\be
\tilde{\chi}_{\rho\rho}(\Bq,\omega)=\int\frac{\dd^3 p}{(2\pi\hbar)^3} f_\FD'(\epsilon_\eq(\Bp)-\mu)\tilde{\nu}(\Bp,\Bq,\omega).\label{chi1}
\ee

\subsection{Quasiparticle tomography}
Plugging Eq.~\eqref{eqn: uext Fourier} into Eq.~\eqref{chgvar} and performing an inverse Fourier transform, we get $\delta f(\Bp,\Br,t)$ as
\be
\delta f(\Bp,\Br,t)=\frac{8\Us}{\pi^2}f_\FD'(\epsilon_\eq(\Bp)-\mu)\times\Re\Big[\tilde \nu_{-}(\Bp)e^{-i(q_\ext y+\ws t)} - \tilde \nu_{+}(\Bp)e^{i(q_\ext y-\ws t)}\Big],
\ee
where $\tilde\nu_\pm(\Bp)\equiv\tilde\nu(\Bp,\pm q_\ext\hat{y},\ws)$. The momentum distribution of the quasiparticles is thus
\be
\delta n(\Bp,t)\equiv\frac{1}{\Vol}\int\diff \Br~ f(\Bp,\Br,t)=\frac{16\Us}{\pi^3}f_\FD'(\epsilon_\eq(\Bp)-\mu)\times\Re\Big[\Big(\tilde \nu_{-}(\Bp) - \tilde \nu_{+}(\Bp)\Big)e^{-i\ws t}\Big].
\ee
Note that $\delta n(\Bp,t)$ is related to $\delta\tilde f(\Bp,\Bq,\omega)$ by
\be
\frac{\pi^3}{16}\frac{\delta n(\Bp,t)}{\Us}=\Re\left[\left(\frac{\delta\tilde{f}(\Bp,-q_\ext\hat{y},\ws)}{\tilde u_\ext(-q_\ext\hat{y},\ws)}-\frac{\delta\tilde{f}(\Bp,q_\ext\hat{y},\ws)}{\tilde u_\ext(q_\ext\hat{y},\ws)}\right)e^{-i\ws t}\right].
\ee
In Fig.~5 of the main text, we show the momentum distributions for the phase $\ws t=\mathcal{N}+\pi/2$, where $\mathcal{N}$ is a positive integer. The integrated momentum distributions in the main text are 
\bea
\delta n_\text{2D}(p_x,p_y)&=&\int\frac{\diff p_z}{(2\pi\hbar)^3}~n(\Bp),\\
\delta n_\text{1D}(p_y)&=&\int\diff p_x~n_\text{2D}(p_x,p_y),
\eea
respectively.

\subsection{Conservation laws}

The collision integral conserves the number, momentum, and total energy of the quasiparticles because the collision process is of the $2\leftrightarrow2$ type. In mathematical form:
\bea
\forall \{\delta{\tilde{f}}(\Bp,\Bq)\},\ \int \frac{\dd^3p}{(2\pi\hbar)^3} \pp^k \tilde{I}_{\rm coll}^{\rm (lin)}(\pp)=0\; \text{ for } k=0,1,2.
\eea
For the kernel of the linearized collision integral Eq.~\eqref{Ncoll}, this leads to 3 zero-energy eigenvectors:
\be
\Gamma(\pp) \pp^k + \int\frac{\diff^3 p'}{(2\pi\hbar)^3} {\pp'}^k \bb{E({\pp,\pp'})-2S({\pp,\pp'})}=0\; \text{ for } k=0,1,2. \label{cons}
\ee

\section{Numerical solutions}
\stepcounter{mycounter}

\subsection{Dimensionless transport equation}
For the numerical solutions, we work in the following (thermal) units
\be
\bar\mu_0=\frac{\mu_0}{\kB T},\quad \bar\omega=\frac{\hbar\omega}{\kB T},\quad \bar \Bq=\frac{\Bq}{k_{T}},\quad \bar \Bp=\frac{\Bp}{p_{T}}, \quad \bas=k_T\as, \quad \bar\chi= \frac{2\kB T}{k_T^3}\tchi_{\rho\rho}=\sqrt{\frac{ T_\F}{T}}\tilde\chi,\label{adim}
\ee
where $p_T=\hbar k_T=mv_T$ is the thermal momentum and $v_T=\sqrt{2\kB T/m}$ is the thermal velocity. Since we are interested in the long-wavelength, low-energy physics (with energies of the order of $\hbar v_T q$), we parametrize the excitation frequency as~\cite{NozieresPines1966SM}
\be
 \bc=\frac{\omega}{\omega_T},
\ee
where $\omega_T=v_T q$. We recall the expression of the collision time
\be
\tau_\coll\equiv\frac{\hbar^3}{2m\as^2}\left(\frac{1}{k_{\rm B}T}\right)^2,
\ee
which is the typical inverse scattering rate of quasiparticles both in the limit $T\ll \TF$ and for $T\approx\TF$.

Eq.~\eqref{limitethermo} can be written in spherical coordinates in dimensionless form as
\begin{multline}
(\bc- p\cos\theta)\tilde{\nu}( p,\theta)-\frac{\bas}{\pi^2} p\cos\theta\int \dd^3 p' g( p')\tilde \nu( p',\theta')+\\
\frac{i}{\omega_T \tau_\coll}\bb{\bar\Gamma(p)\tilde \nu(p,\theta)+\int \dd^3  p' \bbcro{\bar{E}( p, p',u)-2\bar{S}( p, p',u)}\tilde \nu( p',\theta')}=- p\cos\theta,\label{Boltzmannnorm}
\end{multline}
where $\theta$ (resp. $\theta'$) is the angle between $\Bp$ and $\Bq$ (resp. $\Bp'$ and $\Bq$), and $u=\Bp\cdot\Bp'/(pp')=\cos\theta\cos\theta'+\sin\theta\sin\theta'\cos\phi$; here $\phi$ is the azimuthal angle between $\Bp$ and $\Bp'$. For clarity, we dropped the overbars on the dummy (dimensionless) variables $p$ and $p'$, and defined
\bea
g(p')&\equiv&-(\kB T) f_\FD'(\epsilon_\eq(\Bp')-\mu)=\left[4\cosh^2\left(\frac{ p'^2-\bar\mu_0}{2}\right)\right]^{-1}.
\eea

The explicit expressions of the dimensionless kernels $\bar{E}$ and $\bar{S}$ are:
\bea
\bar{E}( p, p',u)&\equiv&\bb{\frac{p_T}{2\pi\hbar}}^3\tau_\coll E(\Bp,\Bp')=\bb{\frac{2\pi\hbar}{p_T}}^3\frac{ k_{\rm B}T}{\pi^4 g^2} E(\Bp,\Bp')=\frac{1}{\pi^2 p_+\sinh\frac{\epsilon+\epsilon'}{2}}\frac{\cosh\frac{\epsilon}{2}}{\cosh\frac{\epsilon'}{2}}\ln\frac{\cosh\frac{\epsilon+\epsilon'+ \bar\mu_0 p_+ p_-}{4}}{\cosh\frac{\epsilon+\epsilon'- \bar\mu_0 p_+ p_-}{4}}\label{E},\\
\bar{S}( p, p',u)&\equiv&\bb{\frac{p_T}{2\pi\hbar}}^3\tau_\coll S(\Bp,\Bp')=\bb{\frac{2\pi\hbar}{p_T}}^3\frac{k_{\rm B}T}{\pi^4 g^2 } S(\Bp,\Bp')=\frac{1}{2\pi^2 p_-\sinh\frac{\epsilon-\epsilon'}{2}}\frac{\cosh\frac{\epsilon}{2}}{\cosh\frac{\epsilon'}{2}}\ln\frac{1+\eee^{-\bar\mu_0({ p_u^2-1})-\epsilon'}}{1+\eee^{-\bar\mu_0({ p_u^2-1})-\epsilon}}\label{S},
\eea
where $\epsilon=p^2-\bar\mu_0$, $\epsilon'=p'^2-\bar\mu_0$, $p_\pm=\sqrt{p^2\pm2pp'u+p'^2}$ and $p_u=p^2p'^2(1-u^2)/p_-^2$. Notice that $\bar{E}( p, p',u)$ (resp. $\bar{S}( p, p',u)$) has a square root divergence when $p=p'$ and $u=-1$ (resp. $u=1$). In the numerical implementation below, we regularize this divergence by using the conservation law Eq.~\eqref{cons}:
\be
\bar\Gamma( p)\equiv\tau_\coll\Gamma(\Bp)=\bb{\frac{2\pi\hbar}{p_T}}^6 \frac{k_{\rm B}T}{\pi^4 g^2}\Gamma(\Bp)=-\int \dd^3 p'(\bar E( p, p',u)-2\bar S( p, p',u)).\label{Gamma} 
\ee

\subsection{Expansion into Legendre polynomials}

To diagonalize the angular part of the collisional kernel Eqs.~\eqref{E}-\eqref{Gamma}, we expand it (as well as the quasiparticle distribution) in the basis of Legendre polynomials:
\bea
\tilde \nu(p,\theta)=\sum_{l=0}^{\infty} \nu_l(p) P_l(\cos\theta),\qquad \nu_l=\int_{0}^\pi \sin\theta\dd\theta \frac{P_l(\cos\theta)}{||P_l||^2}\tilde \nu(p,\theta)\\
\bar E(p,p',u)=\sum_{l=0}^{\infty} E_l(p,p')  \frac{P_l(u)}{||P_l||^2},\qquad E_l=\int_{-1}^1\dd u {P_l(u)}\bar E(p,p',u)
\eea
and similarly for $\bar S$. Here $||P_l||^2=2/(2l+1)$.
The spherical harmonics addition theorem then guarantees that the kernel is diagonal
in the basis of Legendre polynomials:
\be
\int_0^{2\pi}\dd\phi \bar{E}(p,p',u(\theta,\theta',\phi))=2\pi\sum_l E_l \frac{P_l(\cos\theta)P_l(\cos\theta')}{||P_l||^2}.
\ee
Using the orthogonality of the Legendre polynomials and Eq.~\eqref{Gamma}, $\bar{\Gamma}(p)$ can be expressed as
\be
\bar\Gamma(p)=-2\pi\int \diff p'\; p'^2 (E_0(p,p')-2S_0(p,p')). \label{Gammapoly}
\ee
The transport equation Eq.~\eqref{Boltzmannnorm} after decomposing into the Legendre polynomials becomes
\begin{multline}
 \bc \nu_l(p)-p\bb{\frac{l}{2l-1}\nu_{l-1}(p)+\frac{l+1}{2l+3}\nu_{l+1}(p)}-\frac{4}{\pi}\delta_{l1}p\bas \int p'^2\dd p' g( p')\nu_0( p')
\\+\frac{\ii}{\omega_T\tau_\coll}\bb{\bar\Gamma(p)\nu_l(p) +2\pi\int p'^2\dd p'\bb{E_l(p,p')-2S_l(p,p')}\nu_l(p')}=-p\delta_{l1}, \label{nul}
\end{multline}
where $\nu_l(p)=0$ for $l<0$.

\subsection{Numerical implementation}
We calculate $\tilde{\nu}$ by solving numerically the Eq.~\eqref{nul} truncated at $l=l_{\rm max}$. This is a two-step process:
\begin{itemize}
\item Discretization.

We discretize Eq.~\eqref{nul} on a momentum grid $\{p_i\}_{i=1,\cdots,2N_\grid}$, where $N_\grid$ is a large positive integer. 
We separate the momentum grid into two parts: the first part, $p_{1\leq i\leq N_\grid+1}$ forms a Gauss-Legendre quadrature of the interval $[0,p_{\rm m}]$; and the second part $p_{N_\grid+1\leq i\leq2N_\grid}=-\ln\, y_i$ follows from a Gauss-Legendre quadrature $\{y_i\}$ of the interval $[0,\text{exp}(-p_{\rm m})]$, where $p_m=\sqrt{\bar\mu_0}$ is chosen to be where the peak of the density of state is located at low temperatures $T\lesssim\TF$. The discretization of the integrals over $p$ is written as follows:
\be
\int_0^{+\infty} f(p)\dd p  \quad\to\quad \sum_{i=1}^{2N_{\rm grid}} w_i f(p_i)
\ee
where $w_i$ are the weights of the quadratures. For the collision integral in Eq.~\eqref{nul}, this yields 
\be
\bar\Gamma(p_i)\nu_l(p_i) +2\pi\int p'^2\dd p'\bb{E_l(p_i,p')-2S_l(p_i,p')}\nu_l(p')\simeq \sum_{i=1}^{2N_{\rm grid}} M_l(p_i,p_j)\nu_l(p_j)
\ee
where 
\be
M_l(p_i,p_j)=2\pi w_j p_j^2 \int_{-1}^{1}\dd u \bbcro{\bar E(p_i,p_j,u)-2\bar S(p_i,p_j,u)}P_l(u)
\ee
for $i\neq j$, and
\be
M_l(p_i,p_i)=-2\pi\sum_{j\neq i}w_j p_j^2 (E_0(p_i,p_j)-2S_0(p_i,p_j))+2\pi w_i p_i^2 \int_{-1}^{1}\dd u (\bar E(p_i,p_i,u)-2\bar S(p_i,p_i,u))\bbcro{P_l(u)-P_0(u)}
\ee
for $i=j$. Note that we used Eq.~\eqref{Gammapoly} to regularize the integrable divergence in $p_i=p_j$ and $u=\pm1$.

\item Solving the linear equation system.

After the discretization, Eq.~\eqref{nul} becomes a band linear system:
\be
M\vec{\nu}=\vec{s},
\ee
where $\vec{\nu}=(\nu_l(p_i))_{\substack{0\leq i\leq2N_\grid\\ 0\leq l\leq l_{\rm max}}}$ is the unknown vector, $\vec{s}=(-\delta_{l1}p_i)_{(l,i)}$ is the source term
and
\be
M_{(i,l),(j,l')}=\bc \delta_{l,l'}\delta_{i,j}-\delta_{i,j} p_i \bb{\frac{l}{2l-1}\delta_{l',l-1}+\frac{l+1}{2l+3}\delta_{l',l+1}}-\frac{4\bas}{\pi}\delta_{l1}\delta_{l'0}p_i w_j p_j^2 g(p_j)+\frac{\ii}{\omega_T\tau_\coll}\delta_{l,l'} M_l(p_i,p_j)
\ee
is a $2N_\grid( l_{\rm max}+1)\times 2N_\grid( l_{\rm max}+1)$ band matrix with a bandwidth of $2N_\grid+1$. In practice, we use and $N_\grid\geq 300$ and vary $l_{\rm max}$ from $l_{\rm max}=5$ in the hydrodynamic regime, to $l_{\rm max}=1000$ in the collisionless one, where the solution converges as $1/l_{\rm max}$. We verified the convergence with respect to both $N_\grid$ and $l_{\rm max}$. At the end of the calculation, the dimensionless density response function is extracted using Eq.~\eqref{chi1}:
\be
\bar \chi=-\frac{1}{\pi^2}\sum_i w_i p_i^2 g(p_i) \nu_0(p_i).
\ee

\end{itemize}

\section{Solutions in the hydrodynamic and collisionless limits}
\stepcounter{mycounter}
\subsection{Collisionless limit}

\subsubsection{Observability of zero sound in a gas with contact s-wave interactions}
We assess here the conditions for the observability of the zero sound mode in a Fermi gas with weak interactions. This mode disappears on the BCS side ($\kFa<0$) due to the attractive quasiparticle-quasiparticle interactions. It is therefore visible only on the repulsive branch, where metastability limits experimental exploration to $\kFa\lesssim 1$~\cite{ji2022stabilitySM}. We note that a spin zero mode exists on the attractive side $k_\rF a_{\rm s}<0$ according to Mermin's theorem \cite{Mermin1967,stringari2009densitySM}. Although our experiment does not address spin-dependent dynamics, the following discussion on the observability remains valid for the spin mode.

\begin{figure}
\begin{center}
\includegraphics[width=1\textwidth]{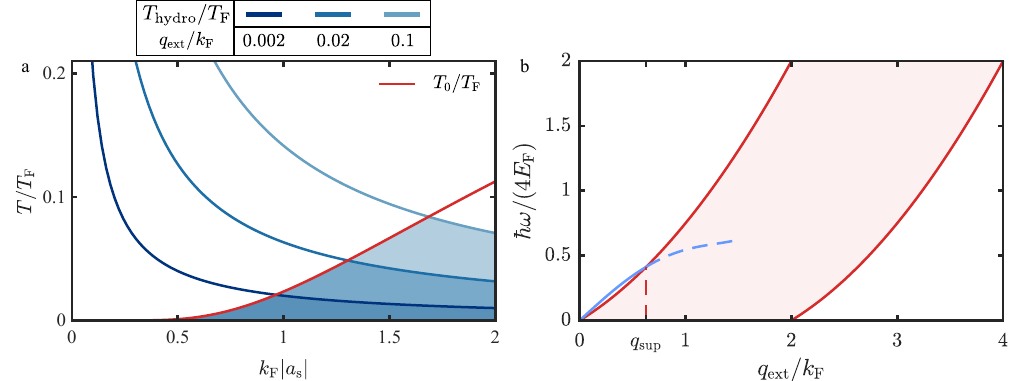}
\caption{\label{observabilite} \textbf{The diagram of observability of zero sound in a Fermi gas with contact interactions.} (\textbf{a}) The zero sound mode is visible only at $T$ below the Landau damping temperature $T_0$ (red curve) and
lower than the transition temperature $T_{\rm hydro}$ to the hydrodynamic regime (blue curves, corresponding to
$\omega_0\tau_{\rm coll}=10$ for various values of $q_\ext/k_\rF$.
(\textbf{b}) The $T=0$ particle-hole continuum and the zero-sound collective branch. For repulsive
interactions ($a>0$), the branch begins as a phononic one with velocity $c>\vF$ and hence is free from damping (at $T=0$).
However, for weak interactions, $c$ is exponentially close to $\vF$, so that the branch enters the continuum at a momentum $q_{\rm sup}$
exponentially close to $0$.}
\end{center}
\end{figure}

In the weakly interacting regime, the zero sound velocity $c_0$ is exponentially close
to $v_\rF$:
\be
 \frac{c_0-v_\rF}{v_\rF} \underset{k_\F |\as|\to0}{\simeq} \frac{2}{\eee^2}\eee^{-\frac{\pi}{k_\F |\as|}} \label{c0intfaible}.
 \ee
This has two consequences. First, this makes the resonance very sensitive to Landau damping. This decay channel, in which the sound wave is absorbed by an energetic quasiparticle, is energetically allowed as soon as the quasiparticle velocity $v_k=\hbar k/m$ is larger than $c_0$. The Landau attenuation rate $\Gamma_{\rm L}$ therefore obeys the activation law
\be
\frac{\Gamma_{\rm L}}{(c_0 -v_\rF)q_\ext}\underset{c_0\to v_\rF}{\simeq}\frac{1}{\pi}{\eee^{-\frac{m( c_0^2-v_\F^2)}{2k_{\rm B}T}}}
\ee
whose activation energy is exponentially suppressed in the weakly interacting regime. Secondly, Eq.~\eqref{c0intfaible} reduces the spectral weight $Z$ of the zero sound by a factor $c_0-v_\rF$. Using as an observability criterion that the maximum  $|Z|/\Gamma_{\rm L}$ of the response function is larger than $1$, we find that $T$ must be below the typical temperature $T_0$:
\be
T_0=\frac{4}{\eee^2}\eee^{-\frac{\pi}{k_\rF|\as|}}\TF \label{T0}
\ee 
This fundamental limitation is depicted by the red curve in Fig.~\ref{observabilite}a. 

If one tries to increase the interaction strength to ease the restriction from Eq.~\eqref{T0}, one encounter the second limitation to the observability of zero sound, namely the onset of the hydrodynamic regime. We consider that zero sound ceases to be observable due to collisional damping as soon as $\omega_0\tau_{\rm coll}<10$, which gives another criterion:
\be
\frac{T_\text{hydro}}{\TF}=\frac{\sqrt{0.2 q_\ext/\kF}}{\kFa}.
\ee
Unlike Eq.~\eqref{T0}, this criterion (depicted by the blue curves in Fig.~\ref{observabilite}a) depends on the wavevector $q_\ext$ at which the sound wave is excited. It is less stringent for larger $q_\ext$, \emph{i.e.}, closer to the collisionless regime. We note however that $q_\ext$ has to remain much smaller than $k_\rF$ (so that we remain in the slowly-varying regime); in the weak coupling limit, we are further constrained to
\be
\frac{q_\ext}{\kF}<\frac{4}{\eee^2}\eee^{-\frac{\pi}{k_\rF|\as|}}.
\ee
This upper bound corresponds to the wavevector at which the sound branch hits the
particle-hole continuum, as depicted in Fig.~\ref{observabilite}b.

Finally, the existence of a superfluid phase below a critical temperature $T_c$ which is of the same order of magnitude as $T_0$ for the attractive gas, poses question as to the very existence of zero sound in a Fermi liquid with purely s-wave interactions.

\subsubsection{Explicit expression of the RPA response function}
Even in the absence of a zero sound resonance, the collisionless response function
significantly deviates from its noninteracting value (the Lindhard function) due to the conservative mean-field force. In the limit $\omega_0\tau_{\rm coll}\gg 1$, the collision integral in Eq.~\eqref{BoltzmannFourier} is negligible, and the remaining integral equation can be solved by a simple resummation, yielding:
\be\label{RPA eq}
\bar\chi_{\rm RPA}{=}\frac{\bar\chi_\rL}{1-4\pi\bas\bar\chi_\rL},
\ee
where we have reused the thermal units Eq.~\eqref{adim} and introduced the Lindhard response $\bar\chi_\rL$ in the long-wavelength limit
\be
\bar\chi_\rL(\bc)=\frac{1}{4\pi^2}f_{\rm L}(\bc)
\ee
where the function $f_{\rm L}$ is given by the Kramers-Kronig relation
\be
f_{\rm L}(\bc)=\int_{-\infty}^{+\infty}\frac{\varrho(\bc')}{ \bc- \bc'}\dd \bc' \quad \text{with} \quad \varrho(\bc)=\frac{\bc}{2}\bb{1-\text{tanh}\frac{ \bc^2-\bar\mu_0}{2}}.
\ee
Note that the exponential behavior of $\varrho(\bc)$ for $\bc>\sqrt{\bar\mu_0}$ and $\bar\mu_0\to+\infty$ is at the origin of the criterion Eq.~\eqref{T0}.

For completeness, we give the fully analytic expression of the $T=0$ Lindhard function in the long-wavelength limit:
\be\label{Lindhard function}
\tilde\chi_\rL(\omega)=\frac{1}{4\pi^2}\bbcro{\bb{\frac{\omega}{\omega_0}~\text{arctanh}\frac{2\, \omega\omega_0}{\omega_0^2+\omega^2}-2}-\ii\pi\frac{\omega}{\omega_0} \Theta\left(\omega_0^2-{\omega}^2\right) }.
\ee

\subsection{Hydrodynamic limit: transport coefficients}\label{sec: hydrodynamic}

Eq.~\eqref{nul} can be solved exactly in the hydrodynamic regime $\omega_T\tau_\coll \ll 1$.
To do so, one can expand the $\nu_l$ on a basis consisting of orthogonal polynomials. Since $\nu_l$ decays exponentially with $l$, $\nu_l \propto (\omega_T \tau_\coll)^{l-1}$, we need only the first three harmonics (in order to calculate the sound attenuation):
\bea
\nu_0(p)&=&\sum_{\substack{n=0 \\ n \text{ even}}}^{+\infty} \nu_n^0 Q_n(p), \label{decnu0}\\
\nu_1(p)&=&\sum_{\substack{n=1 \\ n \text{ odd}}}^{+\infty} \nu_n^1 Q_n(p), \label{decnu1}\\
\nu_2(p)&=&\sum_{\substack{n=2 \\ n \text{ even}}}^{+\infty} \nu_n^2 T_n(p), \label{decnu2}
\eea
where $Q_n$ and $T_n$ are polynomials orthogonal for the same scalar product
\be
\langle Q,T \rangle\equiv\int_{0}^{\infty} p^2 g(p) Q(p) T(p) \dd p
\ee
but initialized differently, \emph{i.e.},
\be
\begin{cases}
\langle Q_n,Q_m \rangle=\delta_{nm}||Q_n||^2, \quad Q_0=1, \quad Q_1=p\\
\langle T_n,T_m \rangle=\delta_{nm}||T_n||^2, \quad \  T_1=p, \quad T_2=p^2
\end{cases}
\ee
where $||Q_n||^2\equiv\langle Q_n,Q_n \rangle$; the coefficients $\nu_n^l$ with $l\geq0$ can be derived using $\nu_n^l=\langle R_n,\nu_l \rangle/||R_n||^2$, where $R_n=Q_n$ for $n=0,1$ and $R_n=T_n$ for $n=2$, respectively. In particular, $\nu_0^0$, $\nu_1^1$, and $\nu_2^0$ are respectively proportional to the density, momentum, and energy, and are thus conserved quantities. Note that $Q_n=T_n$ for $n$ odd. The $T_n$ for $n$ even are divisible by $p^2$, and the decomposition of $\nu_2$ in Eq.~\eqref{decnu2} uses the fact that $\nu_2(0)=0$, due to the continuity of $\nu$ in $\pp=0$ Eq.~\eqref{cons}. Quite generally, the fact that $\nu(\Bp)$ is a differentiable scalar-valued function implies that $\nu_n^l=0$ for $n+l$ odd.

The transport equation \eqref{nul} projected onto $Q_n$ or $T_n$ becomes
\begin{align}
 \bc\nu_n^0&-\frac{1}{3}\bb{\nu^1_{n-1}+\xi_{n+1}^{(Q)}\nu_{n+1}^1}&&&+\frac{\ii}{\omega_T\tau_\coll}\sum_{n'}\mathcal{N}_{nn'}^0 \nu_{n'}^0&=&0&\quad (n\geq0)& \label{nun0}\\
 \bc\nu_n^1&-\bb{\nu^0_{n-1}+\xi_{n+1}^{(Q)}\nu_{n+1}^0}&-&\frac{2}{5}\bb{\nu^2_{n-1}+\xi_{n+1}^{(T)}\nu_{n+1}^2}-\frac{4\bas}{\pi}||Q_0||^2\nu_0^0\delta_{n1}&+\frac{\ii}{\omega_T\tau_\coll}\sum_{n'}\mathcal{N}_{nn'}^1 \nu_{n'}^1&=&-\delta_{n1}&\quad (n\geq1)& \label{nun1}\\
 \bc\nu_n^2&-\frac{2}{3}\bb{\nu^1_{n-1}+\xi_{n+1}^{(T)}\nu_{n+1}^1}&-&\frac{3}{7}\frac{\langle T_n,p\nu_3\rangle}{||T_n||^2}&+\frac{\ii}{\omega_T\tau_\coll}\sum_{n'}\mathcal{N}_{nn'}^2 \nu_{n'}^2&=&0&\quad (n\geq2)& \label{nun2}
\end{align}
with $\xi_{n+1}^{(Q)}=\frac{||Q_{n+1}||^2}{||Q_{n}||^2}$, $\xi_{n+1}^{(T)}=\frac{||T_{n+1}||^2}{||T_{n}||^2}$, $\nu_n^l=0$ for $n<0$, and
\begin{multline}
\mathcal{N}_{nn'}^{l}=\int_0^{+\infty}p^2\dd p g(p)  \bar \Gamma(p) \frac{R_n(p)R_{n'}(p)}{||R_n||^2}+2\pi\int_0^{+\infty}p^2\dd p g(p)p'^2\dd p' \frac{R_n(p)}{||R_n||^2}\bb{E_l(p,p')-2S_l(p,p')}R_{n'}(p')\\
=2\pi\int_0^{+\infty}p^2\dd p g(p)p'^2\dd p' \int_{-1}^{1}\dd u\frac{R_n(p)}{||R_n||^2}\bbcro{\bar E(p,p',u)-2\bar S(p,p',u)}\bbcro{R_{n'}(p')P_l(u)-R_{n'}(p)P_0(u)}
\end{multline}
where $R_n=Q_n$ for $n=0$ or $1$ and $R_n=T_n$ for $n=2$ and we have reexpressed $\bar{\Gamma}(p)$ using the conservation law Eq.~\eqref{Gamma} in the second line. Furthermore, Eq.~\eqref{Gamma} turns into $\mathcal{N}_{n0}^0=\mathcal{N}_{0n'}^0=0$ for the total number of quasiparticles, $\mathcal{N}_{n2}^0=\mathcal{N}_{2n'}^0=0$ for the energy and $\mathcal{N}_{n1}^1=\mathcal{N}_{1n'}^1=0$ for the momentum. As a consequence, the three conserved quantities $\nu_0^0$, $\nu_2^0$ and $\nu_1^1$ are of order unity in the hydrodynamic limit, and the non-conserved quantities directly coupled to them are subleading in $\omega_T\tau_\coll$. Written as vectors, they are
\bea
\vec{\nu}^2&\equiv&(\nu_n^2)_n=\mathcal{O}(\omega_T\tau_\coll) \label{vecnu2}\\
\vec{\nu}^{1,\perp} &\equiv&(\nu_{n>1}^1)_n=\mathcal{O}(\omega_T\tau_\coll). \label{vecnu1}
\eea 
All the other coefficients (in particular the $\nu_n^3$ and $\nu_{n>2}^0$) are at least of order $\mathcal{O}(\omega_T\tau_\coll)^2$ and therefore negligible to the order at which we solve the system Eqs.~\eqref{nun0}-\eqref{nun2}.

In particular, the equations for the conserved quantities can be derived by setting $n=0,2$ in Eq.~\eqref{nun0} and $n=1$ in Eq.~\eqref{nun1}:
\bea
\bc\nu_0^0-\frac{1}{3}\xi_{1}^{(Q)}\nu_{1}^1&=&0\label{nun0cons1}\\
\bc\nu_2^0-\frac{1}{3}\bb{\nu^1_{1}+\xi_{3}^{(Q)}\nu_{3}^1}&=&0\label{nun0cons2}\\
\bc\nu_1^1-\bb{\nu^0_{0}+\xi_{2}^{(Q)}\nu_{2}^0}-\frac{2}{5}\xi_{2}^{(T)}\nu_{2}^2-\frac{4\bas}{\pi}||Q_0||^2\nu_0^0 &=&-1\label{nun1cons1}.
\eea
The transport equations for the non-conserved quantities can be derived by writing Eq.~\eqref{nun2} and Eq.~\eqref{nun1} ($n\geq2$) in matrix form
\bea
\frac{i}{\omega_T\tau_\coll}\mathcal{N}^2\vec{\nu}^2 &=&\frac{2}{3}\nu_1^1\hat{u}_2\label{nu2eqn}\\
\frac{i}{\omega_T\tau_\coll}\mathcal{N}^{1,\perp}\vec{\nu}^{1,\perp} &=&{\nu_2^0\hat{u}_3},\label{nu1perpeqn}
\eea
where the unit vectors $\hat{u}_k=(\delta_{kn})_n$, $\mathcal{N}^2=(\mathcal{N}^2_{nn'})$ and $\mathcal{N}^{1,\perp}=(\mathcal{N}^1_{n>1,n'>1})$.
From Eqs.~\eqref{nu2eqn}-\eqref{nu1perpeqn}, one can express $\nu_2^0$ and $\nu_3^1$ in terms of the conserved quantities:
\bea
\nu_2^2&=&{-\frac{2i}{3}\omega_T \tau_\eta}\nu_1^1 \label{nu20}\\ 
\nu_3^1&=&{-{i}\omega_T \tau_\kappa}\nu_2^0, \label{nu31}
\eea
where we have introduced two relaxation times $\tau_\eta$ and $\tau_\kappa$:
\begin{eqnarray}
\frac{\tau_\eta}{\tau_\coll}&=&\hat{u}_2 \frac{1}{\mathcal{N}^2}\hat{u}_2  \\ 
\frac{\tau_\kappa}{\tau_\coll} &=&\hat{u}_3 \frac{1}{\mathcal{N}^{1,\perp}}\hat{u}_3.
\end{eqnarray}
We substitute the expressions Eqs.~\eqref{nu20}-\eqref{nu31} into the transport equation of the conserved quantities Eqs.~\eqref{nun0cons1}-\eqref{nun1cons1}. Solving the remaining 3 by 3 linear system for the density response function, we find
\be
\bar\chi=-\frac{1}{\pi^2}\nu_0^0||Q_0||^2=-\frac{1}{\pi^2}\frac{||Q_1||^2/3}{\bc^2-\bc_1^2+2i\bc{  \gamma(\bc)\omega_T\tau_\coll}},
\ee
where the first sound velocity is
\be\label{c1}
\bc_1=\sqrt{\frac{\xi_1^{(Q)}(1+4\bas ||Q_0||^2/\pi)+\xi_2^{(Q)}}{3}}
\ee
and the sound attenuation coefficient
\be
\gamma(\bc)={\frac{\xi_2^{(Q)}\xi_3^{(Q)}}{18\bc^2}\frac{\tau_\kappa}{\tau_\coll}+\frac{2\xi_2^{(T)}}{15} \frac{\tau_\eta}{\tau_\coll}}.
\label{gammac}
\ee
The complex pole of $\bar\chi(\bc)$ is located at $\bc=\bc_1-i\gamma(\bc_1)\omega_T\tau_\coll$.
Going back to the physical units, the sound attenuation coefficient $\Gamma(\bc_1)$ is expressed in terms of the shear viscosity $\eta$ and thermal diffusivity $D_\kappa$
\be
\Gamma(\bc_1)\equiv2\gamma(\bc_1)\omega_T^2\tau_\coll=\frac{2}{3}\frac{\eta q^2}{m\rho_\eq}+\frac{\xi_2^{(Q)}}{6\bc_1^2} D_\kappa q^2
\label{Gammac1}
\ee
and
\bea
\frac{\eta}{m\rho_\eq}&=&\frac{2\xi_2^{(T)}}{5}v_T^2 \tau_\eta\\
D_\kappa&=&\frac{\xi_3^{(Q)}}{3}v_T^2\tau_\kappa.
\eea
The dimensionless coefficients $\xi_n$ can be calculated explicitly using the equation of state of the ideal gas; they can be expressed in terms of Fermi-Dirac
 integrals by noting that $\langle p^k,p^n\rangle=\Gamma(\frac{n+k+1}{2})\mathcal{F}\bb{\frac{n+k-1}{2}}$. We note that the bulk viscosity does not contribute to the sound attenuation Eq.~\eqref{Gammac1} in this weakly interacting regime. This is due to the parabolic dispersion relation, $\partial\epsilon/\partial\Bp\propto \Bp$, which couples the total velocity $\nu_1^1$ only to the conserved quantities $\nu^0_0$ and $\nu^0_2$ [see Eq.~\eqref{nun1cons1}] and not to the dissipative isotropic components $\nu^0_{n>2}$.

The sound attenuation coefficient is often calculated using approximate solutions of the transport equation. The most common (and most reliable one) is the so-called variational (or relaxation time approximation)
\cite{Smith2005bSM,Smith2005SM,PhysRevA.82.033619SM},
which neglects the energy dependence of the quasiparticle distribution, retains only the first term in the expansions Eq.~\eqref{decnu0}-\eqref{decnu2}, and therefore approximates the relaxation times and attenuation coefficient by
\begin{eqnarray}
\frac{\tau_\eta^{(0)}}{\tau_\coll}&=& \frac{1}{\mathcal{N}^2_{22}}, \\ 
\frac{\tau_\kappa^{(0)}}{\tau_\coll} &=&\frac{1}{\mathcal{N}^{1,\perp}_{33}}, \\
\gamma^{(0)}&=&{\frac{\xi_2^{(Q)}\xi_3^{(Q)}}{18\bc^2}\frac{\tau_\kappa^{(0)}}{\tau_\coll}+\frac{2\xi_2^{(T)}}{15} \frac{\tau_\eta^{(0)}}{\tau_\coll}}.  \label{gamma0}
\end{eqnarray}
$\gamma^{(0)}$ underestimates the sound attenuation with an error $|\gamma-\gamma^{(0)}|/\gamma^{(0)}$ of about 8\% in the low-$T$ regime.

Note that the terminology ``relaxation time approximation''
is somewhat ambiguous as it may also refer to the rougher approximation used by Abrikosov and Khalatnikov~\cite{khalatnikov1958dispersionSM}, which was to set $p=p_\F$ and $T/\TF\ll1$ in Eq.~\eqref{nul}~\cite{dispersionSM}; this approximation overestimates the sound attenuation by $\approx 20\%$ [see the inset of Fig.~4(b) in the main text].

\subsection{Crossover from the zero sound to the first sound}

\begin{figure*}[!hbt]
  \centering
  \includegraphics[width=1\columnwidth]{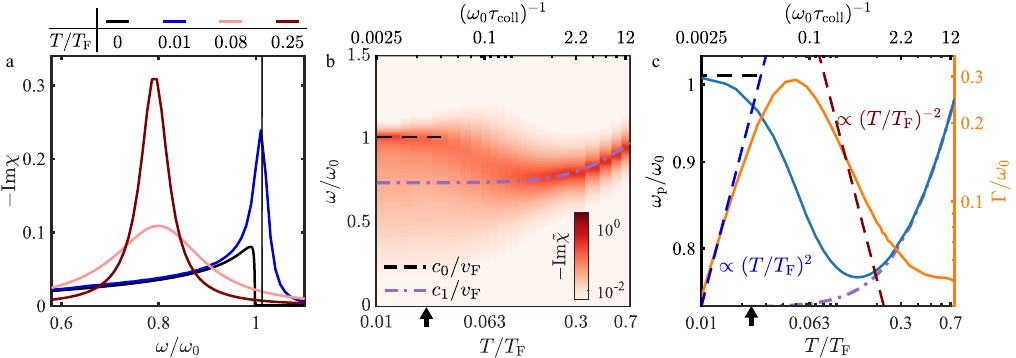}
  \caption{ \textbf{Crossover from the zero sound to the first sound in a repulsive Fermi gas with contact interactions.} (\textbf{a}) $-\Im\tchi$ versus $\omega/\omega_0$ for various $T/\TF$. The colors correspond to different temperatures (see legend). (\textbf{b}) Heatmap of $-\Im\tchi$ at different $T/\TF$ (lower axis) and the corresponding $(\omega_0\tau_\coll)^{-1}$ (upper axis). The color bar shows the magnitude of $-\Im\tchi$ in a logarithmic scale. The black dashed line is the normalized zero sound velocity $c_0/\vF\approx1.013$, and the purple dot-dashed line is the normalized first sound velocity $c_1/\vF$ calculated from Eq.~\eqref{c1}. The black arrow shows $T_0/\TF$.  (\textbf{c}) The peak frequency $\omega_\p/\omega_0$ (blue solid line, left axis) and the width $\Gamma/\omega_0$ (orange solid line, right axis) extracted from the spectra shown in (b) as a function of $T/\TF$ (lower axis) and the corresponding $(\omega_0\tau_\coll)^{-1}$ (upper axis). The black dashed line and the purple dot-dashed line have the same meaning as in (b). The blue and red dashed lines are the references for the collisional damping rate in the collisionless regime $\Gamma\propto(T/\TF)^{2}$ and in the hydrodynamic regime $\Gamma\propto(T/\TF)^{-2}$, respectively. For this figure, the gas parameter is $\kFa=1$ and the sound wavevector is $q_\ext/\kF=0.02$.
  \label{FIG:crossover}}
\end{figure*}

In this section, we discuss the (smooth) transition from zero to first sound in the Fermi gas with repulsive contact interactions. In Fig.~\ref{FIG:crossover}, we show this crossover for the gas parameter $\kFa=1$ and sound wavevector $q_\ext/k_\F=0.02$. As discussed previously, the zero sound mode exists in the collisionless regime ($\omega_0\tau_\coll\gg1$) and the first sound exists in the hydrodynamic regime ($\omega_0\tau_\coll\ll1$). In Fig.~\ref{FIG:crossover}a, we show $\Im\tchi$ for various $T/\TF$, corresponding to different collision rates. At $T=0$, $\tchi$ is calculated from Eq.~\eqref{RPA eq} and Eq.~\eqref{Lindhard function}: it consists of a broad Lindhard-like spectrum for $\omega/\omega_0\leq1$ and a Dirac delta function located at $\omega=c_0 q_\ext$.

As $T$ increases towards $0.01\TF$, the Dirac peak merges with the broad spectrum, resulting in a sharp peak at $\omega\approx c_0 q_\ext$. This peak gets broadened as $T$ increases due to the collisional damping, until $T\gtrsim T_0$ (see Fig.~\ref{observabilite}a). Finally, the system enters the hydrodynamic regime at $T\gtrsim0.2\TF$ and the first sound appears as the dominant collective excitation. Fig.~\ref{FIG:crossover}b is the heatmap of the spectra $\Im\tchi$ for different $T/\TF$. The black arrow shows $T_0/\TF$ (and $T_\text{hydro}/\TF\approx0.063$ is located at $\omega_0\tau_\coll=10$). 

The narrow peaks at low and high $T$ thus correspond to the zero sound and the first sound, respectively; in the crossover regime, the spectra are broad. The black dashed line is the normalized zero sound velocity $c_0/\vF\approx1.013$, and the purple dot-dashed line shows the normalized first sound velocity $c_1/\vF$ calculated from Eq.~\eqref{c1}. In Fig.~\ref{FIG:crossover}c, we show the peak frequency (blue solid line, left axis) and the full width at half maximum (orange solid line, right axis) as a function of $T/T_\F$ extracted from the spectra shown in Fig.~\ref{FIG:crossover}b. The blue and the red dashed lines are the references for the collisional damping in the collisionless regime $\Gamma/\omega_0\propto (T/\TF)^{2}$ for the zero sound and $\Gamma/\omega_0\propto(T/\TF)^{-2}$ in the hydrodynamic regime for the first sound.

\section{Verifying the linearity of the density response}
\begin{figure*}[!hbt]
  \centering
  \includegraphics[width=1\columnwidth]{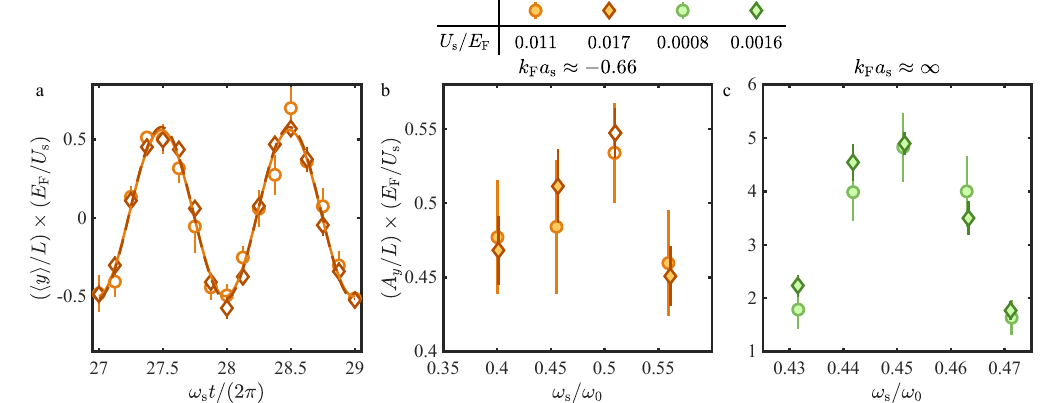}
  
  \caption{\textbf{Verifying the linearity of the density response.} (\textbf{a}) The time-resolved measurement of the normalized CoM for $\omega_\s/\omega_0\approx0.5$ and $\kFa\approx-0.66$ with two drive strength $\Us/\EF\approx0.011$ (circles) and $\Us/\EF\approx0.017$ (diamonds). The dashed and solid lines are the sinusoidal fits on the data. (\textbf{b}) and (\textbf{c}) are the normalized amplitude of the CoM oscillation for $\kFa\approx-0.66$ and $\kFa\approx\infty$, respectively. The corresponding $\Us/\EF$ for different data sets are shown in the legend.
  \label{FIG:LinearResponse}}
\end{figure*}

\stepcounter{mycounter}
According to Eq.~\eqref{CoMy LR}, the signature of linear response is that $\braket{y}\propto U_\s$. Experimentally, we adjust $\Us$ and verify that $\braket{y}/U_\s$ is constant for each $\ws$. We show examples of such measurements in Fig.~\ref{FIG:LinearResponse} for $\kFa\approx-0.66$ and $\kFa\approx\infty$. In Fig.~\ref{FIG:LinearResponse}a we plot the normalized CoM $\braket{y}/L\times \EF/U_\s$ at $\Us/\EF\approx0.011$ (circles) and $0.017$ (diamonds) for $\omega_\s/\omega_0\approx0.5$, (which corresponds to the peak response of $\Im\tchi$ at $\kFa\approx-0.66$). The two traces collapse onto the same curve, showing that the system is indeed in the linear response regime. In Fig.~\ref{FIG:LinearResponse}b and c we show $A_y/L\times\EF/\Us$ extracted by a sinusoidal fit on the normalized CoM for $\kFa\approx-0.66$ and $\kFa\approx\infty$  respectively. The good overlaps between these data sets again confirm that the response of the system is linear in the drive.

\section{Fits to the density response functions}
\begin{figure*}[!hbt]
  \centering
  \includegraphics[width=1\columnwidth]{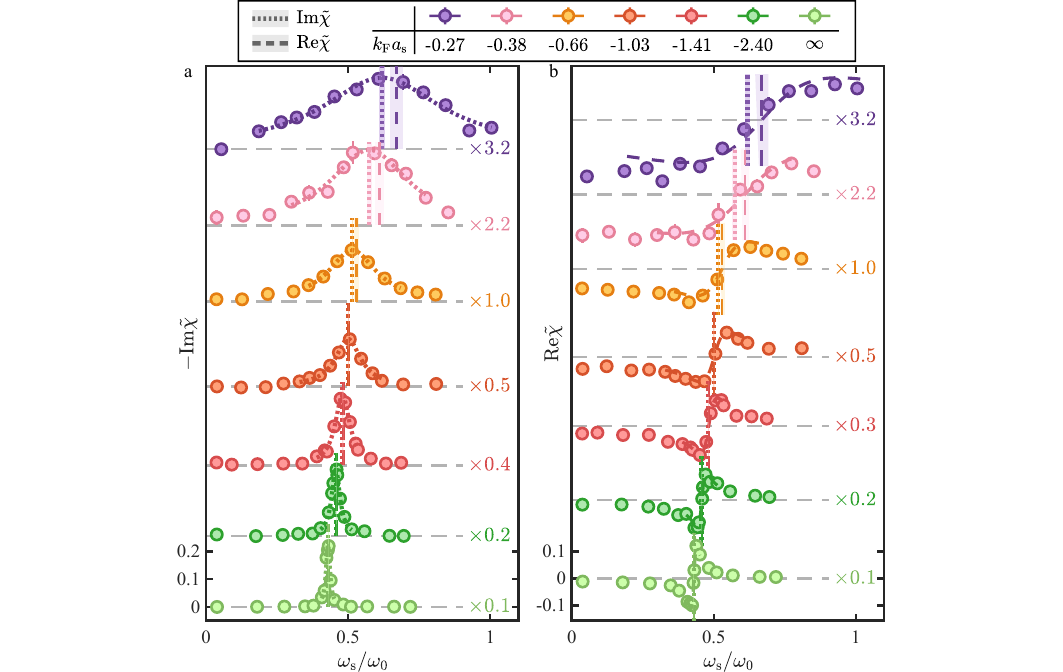}
  
  \caption{\textbf{Fits to the density response functions.} We show the fits from which we extract the data of Fig.~2c-d of the main text. The dotted (resp. dashed) lines are the fits for $\Im\tchi$ (resp. $\Re\tchi$); the extracted $\omega_\p/\omega_0$ is shown as the vertical dotted (resp. dashed) lines. The bands are the uncertainties of the fit.
  \label{FIG:fit}}
\end{figure*}

\stepcounter{mycounter}
In Fig.~\ref{FIG:fit}, we show the fits for the spectra of Fig.~2c-d of the main text. The explicit expressions of the fitting functions are 
\begin{align}
    \Im\tchi_\text{fit}(\omega_\s)&=-\frac{Z}{\pi}\frac{\Gamma/2}{(\omega_\s-\omega_\text{p})^2+(\Gamma/2)^2},\\
    \Re\tchi_\text{fit}(\omega_\s)&=\frac{Z}{\pi}\frac{\omega_\s-\omega_p}{(\omega_\s-\omega_\text{p})^2+(\Gamma/2)^2}.
\end{align}
We constrain the fitted region so that $|\Im\tchi|<0.2\text{Max}(|\Im\tchi|)$ for $|\kFa|<1$ and $|\Im\tchi|<0.1\text{Max}(|\Im\tchi|)$ for $|\kFa|>1$. We checked that varying this range does not affect the results of Fig.3 of the main text.
The fitted spectra are shown as the dotted (dashed) lines for the $\Im\tchi$ ($\Re\tchi$) in Fig.~\ref{FIG:fit}a (b), with $\omega_\p/\omega_0$ marked as the dotted (dashed) lines. The bands are the uncertainties of the fit.

\begin{thebibliography}{69}
\bibitem [{dyn()}]{dynamicinst}%
  \BibitemOpen
  \href@noop {} {}\bibinfo {note} {Note that the first-order theory also exhibits a dynamical instability for $\kFa<-\pi/2$, which is not seen experimentally, further showing its limitations.}\BibitemShut {Stop}%
\bibitem [{sou()}]{soundspeed}%
  \BibitemOpen
  \href@noop {} {}\bibinfo {note} {In general, $c_{1,T}/c_1=\sqrt{C_V/C_P}\leq 1$, where $C_V$ (resp. $C_P$) is the heat capacity at constant volume (resp. pressure). At $T=0$, $c_{1,T}=c_1$.}\BibitemShut {Stop}%
\bibitem [{exa()}]{exactres}%
  \BibitemOpen
  \href@noop {} {}\bibinfo {note} {The exact result is $a=\frac{\pi}{45}+\frac{4\alpha}{15\pi}\sum_{n=0}^\infty \frac{2n+1}{n^2(n+1)^2}\frac{1}{n(n+1)-\alpha}\approx 0.086$ for $\alpha=2/5$.}\BibitemShut {Stop}%%%%
\end{thebibliography}
\end{document}